\documentclass[aps,prl,groupedaddress,twocolumn]{revtex4-2}
\usepackage{graphicx}
\usepackage{xcolor}
\usepackage{amsmath}
\usepackage{amssymb}
\usepackage{bm}
\usepackage{ulem}
\usepackage{svg}
\DeclareMathOperator*{\argmax}{arg\,max}
\begin{document}
\title{Quantum readout of imperfect classical data}
\author{Giuseppe Ortolano$^{1,2}$}
\author{Ivano Ruo-Berchera$^{1,*}$}
\affiliation{$^{1}$Istituto Nazionale di Ricerca Metrologica, strada delle Cacce, 10135-Torino, Italy\\
$^{2}$DISAT, Politecnico di Torino, Corso Duca degli Abruzzi 24, 10129 Torino, Italy}
\affiliation{$^*$ email: i.ruoberchera@inrim.it}

\begin{abstract}
The encoding of classical data in a physical support can be done up to some level of accuracy due to errors and the imperfection of the writing process. Moreover, some degradation of the storage data can happen over the time because of physical or chemical instability of the system. Any read-out strategy should take into account this natural degree of uncertainty and minimize its effect.
An example are optical digital memories, where the information is encoded in two values of reflectance of a collection of cells. Quantum reading using entanglement, has been shown to enhances the readout of an ideal optical memory, where the two level are perfectly characterized. In this work, we will analyse the case of imperfect construction of the memory and propose an optimized quantum sensing protocol to maximize the readout accuracy in presence of imprecise writing. The proposed strategy is feasible with current technology and is relatively robust to detection and optical losses. Beside optical memories, this work have implications for identification of pattern in biological system, in spectrophotometry, and whenever the information can be extracted from a transmission/reflection optical measurement.
\end{abstract}
\maketitle
\section{Introduction}

The use of quantum resources, such as quantum correlations and squeezing, has allowed to surpass classically imposed limits on a variety of practical tasks. Restricting to the optical domain, quantum metrology and sensing \cite{Giovannetti_2011,Pirandola_2018,Berchera_2019} shows the possibility to improve parameter estimation\cite{Giovannetti_2004}, such as phase \cite{Schnabel_2017,Schafermeier_2018,Ortolano_2019} and transmission \cite{Monras_2007,Adesso_2009,Losero_2018}, with relevant applications both to technology \cite{Brida_2010,Genovese_2016} and fundamental physics \cite{Aasi_2013,Berchera_2013,Pradyumna_2020}. In quantum hypothesis testing \cite{Helstrom_1976,Chefles_1998}, a certain number of protocols have been proposed, in particular the quantum illumination \cite{Lloyd_2008,Tan_2008,Lopaeva_2013,Zhang_2020,Gregory_2020} addressed to the target detection in noisy background and the quantum reading (QR) \cite{Pirandola_2011,Nair_2011,Dallarno_2012,Wilde_2012,Dallarno_2012a,Hirota_2017,Pereira_2022}, capable of improving the readout of data stored in classical digital memories. 
In classical memories information is encoded in a physical object for later reuse, and a successful encoding also depends on the reliability of the writing (encoding) process. An example are optical digital memories, where the information is encoded in two possible values of the reflectance (or equivalently transittance) of a collection of cells. In this context, the QR  protocol\cite{Pirandola_2011} shows a significant quantum enhancement of the read-out performance, in terms of bits extracted from a cell for a fixed energy. Single cell QR has been recently realized experimentally \cite{Ortolano_2021} and has been theoretically generalized to a more realistic multicell scenario \cite{Pirandola_2011a,Zhuang_2020,Oskuei_2021,Revson_2021}. While those limits are important in gauging the possible improvement offered by quantum resources over classical strategies, in a more application oriented approach, it is useful to consider the case when the transmittance values cannot be reproduced with arbitrary precision, rather they can be more realistically represented by classical random variables whose distributions can be eventually characterized. The scenario could be the one of a commercial production with a limited single-cell accuracy  but with the possibility of an extremely precise post-production characterization. In this work, we analyse the effect of possible defects in the construction of the memory on the readout performance and maximize the readout accuracy in presence of an imperfect, yet characterized, writing process. 

The problem studied here has some analogy to the recent proposed task of quantum conformance test \cite{Ortolano2021a}, although the goal is different, the last one being devoted to the recognition of a defective production process with respect to a standard.
Moreover, we extend the analysis of QR with imperfect cells by considering a more general classical benchmark that takes into account a multicell memory and a collective measurement of the probed cells, showing that cell-by-cell quantum readout remains anyway better in recovering the stored information. A multicell memory can be seen as a large block of cells for which the information is stored, according to some classical encoding, in classical codewords expressed by cells (quantum channels). In the limit of very large memories (infinte number of cells) the maximum amount of information that can be retrieved from the memory can by found with a constrained (at fixed energy) optimization of the Holevo bound \cite{Holevo_1973,Holevo_1998,Hausladen_1996} that we solve in the case of classical input states and imperfect encoding.

\section{Materials and Methods}

\subsection*{Optical memory and readout model}
\begin{figure}\label{scheme}
	\vspace{0.10cm}
	\center
	\includegraphics[width=\columnwidth]{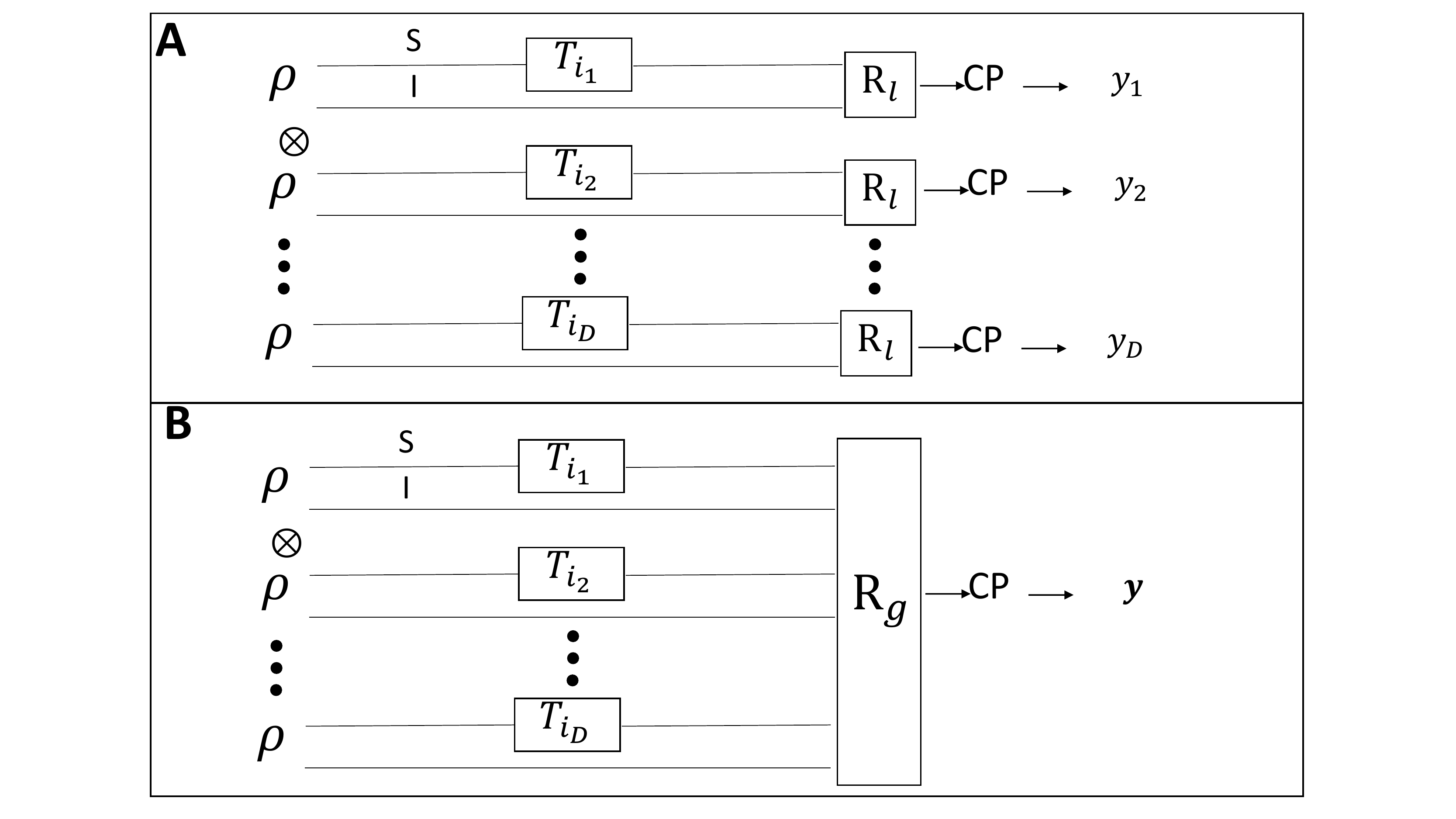}
	\caption{\label{fig:scheme}\textbf{Memory readout scheme.} \textbf{A.} \textit{Single cell independent reading.} A bipartite state $\rho$ consisting of an $M$ modes signal system and a $L$ modes idler one, is irradiated by a transmitter. The signal interacts with the memory cell whose transmittance is a random variable $T_i$, $i=0,1$, depending on the value of the bit. The idler is sent directly to a local receiver ($R_l$) where it is measured jointly with the signal. The value of the bit is determined by classical post-processing of the measurement result. In the case of perfect memory the random variable $T_i$ reduce to a single parameter $\tau_i$. \textit{Parallel reading of a memory.} If an array of length $D$ (a memory) is tested n parallel, instead of a single cell, $D$ copies of $\rho$ are sent to the cells, one for each, and a joint measurement between all the copies is performed at the receiver, this time a global one ($R_g$).}
\end{figure}

Let us consider an optical memory cell storing one bit of information in two possible values of trasmittance $\tau_0$ and $\tau_1$. The readout of the cell is carried out using a transmitter emitting a bipartite optical probe in a state $\rho$. Let us call the two systems of the bipartite state signal (S) and idler (I) system. A number $M$ of modes in the signal system are sent to the cell while $L$ modes of the idler system are sent directly to a receiver where a general joint POVM with the returning signal is performed. A decision on the value, $y=0,1$, of the stored bit is taken after classical post processing of the measurement result. A schematic of the protocol is given in Fig.\ref{fig:scheme}A. The optical transmittance acting on the input state can be modelled by means of a pure quantum loss channel $\mathcal{E}_{\tau}$, so that the state at the receiver is written as $\sigma=(\mathcal{E}_{\tau}\otimes\mathcal{I})\rho$, where $\mathcal{I}$ is the identity. The problem of information recovery can then be seen as a problem of quantum channel discrimination \cite{Pirandola_2019,Zhuang_2020a} between two channels, $\mathcal{E}_{\tau_{0}}$ and $\mathcal{E}_{\tau_{1}}$, with the minimum probability of error. When the discrimination is performed using optical states that lives in an infinite dimensional Hilbert space, a non trivial formulation of the problem requires that some energetic constraint are imposed on the states \cite{Holevo_2012}. A common choice, that we adopt here, is to fix the energy of the signal system. The minimization of the probability of error with an energy constraint requires a double optimization, both on the input state and on the  measurement performed at the receiver. In case of a perfectly characterized memory, for which $\tau_0$ and $\tau_1$ are known with arbitrary accuracy, ultimate theoretical limits on the readout performance has been found \cite{Pirandola_2011,Pirandola_2011a,Nair_2011}.
A single cell of an imperfect optical memory stores one bit of information in one of two possible random values of transmittance, $T_{i}$, with $i=0,1$, each with a probability distribution $g_{i}$. The information is recovered, in general, with the same procedure described in the previous section for a perfect memory cell, that is then a particular case of the imperfect cell, for which $T_0$ and $T_1$ take only the real values $\tau_0$ and $\tau_1$ respectively. The problem of information retrieval from an imperfect memory cell remains formally similar to the perfect memory one, namely it remains a quantum channel discrimination problem, but the channel to be discriminated are convex combinations of pure loss channels $\mathcal{E}_\tau$. In particular a bipartite input state $\rho$ irradiated by the transmitter will be traced to the state $\rho_i$:
\begin{equation}
\rho_i=\mathbb{E}_{g_i}[(\mathcal{E}_{\tau} \otimes \mathcal{I})\rho] \label{channel}
\end{equation}
where $i=0,1$ and $\mathbb{E}_{g_i}[\cdot]$ denotes the expectation value over the distribution $g_i$.
The problem of discriminating two channels in the form of Eq.(\ref{channel}) has been analysed in a different context in the Quantum Conformance Test (QCT) protocol \cite{Ortolano2021a}. In particular it was shown that, for a fixed number of signal photons $\mu$, classical states, i.e. convex combinations of coherent states of the form:
\begin{equation}
\rho^{cla} = \int d^{\text{\tiny$2M$}}\bm{\alpha}\, d^{\text{\tiny$2L$}} \bm{\beta} \, P\left(\bm{\alpha}, \bm{\beta} \right)|\bm{\alpha}\rangle \langle \bm{\alpha}| \otimes |\bm{\beta}\rangle \langle \bm{\beta}|
\label{cla}
\end{equation}
have probability of error in the discrimination that is lower bounded by:
\begin{equation}
p^{cla}_{err} \geq \frac{ 1- \mathbb{E}_{g_0}\left[\mathbb{E}_{g_1}\left[\sqrt{1-e^{-\mu\left(\sqrt{\tau_0} - \sqrt{\tau_1 }\right)^2}}\right]\right]}{2}
\label{hb}
\end{equation}
for any input state and output measurement. The use non classical states, in particular two mode squeezed vacuum states \cite{Braunstein_2005,Weedbrook_2012} paired with photon counting measurements allows to surpass the classical limit in Eq.(\ref{hb}). 

In the context of memory reading a more fitting figure of merit is the information recovered by the procedure. We consider the cell prepared with equal probability in $T_0$ or $T_1$, namely $p_0=p_1=1/2$. Given the probability of error $p_{err}$ the information recovered $I$ will be:
\begin{equation}
I=1-H(p_{err})
\end{equation}
where $H(p)=-p \log_2 p -(1-p)\log_2 (1-p)$ is the binary Shannon entropy.
In the following, we will compare the performance of a specific 'local' quantum read-out strategy, where local means that each cell is probed and measured separately, with respect to three different classical benchmarks: The local optimal one, the local based on a specific receiver (photon counting), and finally a 'global' optimal one. In the last one, the cells are still probed one by one, but the receiver is allowed to perform a collective measurement across the memory, as it is depicted in Fig.\ref{fig:scheme}B.

\subsection*{Local Classical limits.}
Consider the scheme in Fig.\ref{fig:scheme}A, and a classical input state as defined in Eq.(\ref{cla}). If a single cell is probed the probability of error is lower bounded by the expression in Eq.(\ref{hb}), meaning that the information recovered will be upper bounded by:
\begin{equation}
\mathcal{C}^{HB}:=1-H(p^{cla}_{err}) \label{CHB}
\end{equation}
where we chose the superscript $HB$ to denote this limit since its derivation is based on the Helstrom bound \cite{Helstrom_1976}.

The limit in Eq.(\ref{CHB}) refers to an optimal unspecified detection scheme. Because of certain number of inequality used to recover it, this lower bound is expected to be not tight, meaning that it could not be reachable. Thus, it is useful to consider a second local classical benchmark choosing a specific receiver, in particular a photon counting receiver, which is close to the optimality for transmission estimation when paired with a maximum likelihood post-processing decision procedure. A detailed discussion on this limit is given in Ref.\cite{Ortolano2021a}. Let's denote the probability of error of this scheme as $p_{err}^{cla,PC}$. We can define a second informational limit as:
\begin{equation}
\mathcal{C}^{PC}:=1-H(p^{cla,PC}_{err}) \label{CPC}
\end{equation}
The bound in Eq. (\ref{CPC}) represents the maximum information that can be extracted from the memory assuming a classical input state and a photon counting measurement at the receiver.

\subsection*{Local quantum strategy.}
Let us consider now input states not belonging to the class defined by Eq.(\ref{cla}), i.e. non-classical state. In particular we choose a collection $\rho=\bigotimes^{M}\vert TMSV\rangle_{I,S}$, of $M$ Two Mode Squeezed Vacuum (TMSV) states, represented in the Fock base $\{|n\rangle\}$  as:
\begin{equation}
\vert TMSV\rangle_{I,S}\propto\sum_{n} \sqrt{P_{\mu}(n)}\vert n\rangle_{P}\vert n\rangle_{R},
\end{equation}
where $P_{\mu}(n)=\mu^n/(\mu + 1)^{n+1}$ is a thermal distribution with mean photon $\mu$. It is a bipartite maximally entangled state with perfectly correlated number of photons of the signal and idler systems \cite{Bondani_2007,Avella_2016,Meda_2017}. We have in this case $L=M$. After the interaction with the memory cell the receiver performs a photon counting measurement both at the signal and idler systems. Then, a maximum likelihood processing is applied to each pair of data, to extract the bit value. The value  $y$ assigned to the recovered bit -- given the conditional probability $p\left(T_i|\bm{n}\right)$ of the bit having value $i=0,1$ after measuring $\textbf{n}=(n_S,n_I)$ photons for signal and idler system at the receiver -- is given using the condition $y=\argmax_{i=0,1}p\left(T_i|\bm{n}\right)$. This choice, given the constant prior, $p(T_0)=p(T_1)=1/2$,  is equivalent to a maximum likelihood decision, i.e. $y=\argmax_{i=0,1}p\left(\bm{n}|T_i\right)$

The steps for the derivation of the probability of error, $p^{qua,PC}_{err}$, can be fount in  \cite{Ortolano2021a}, and we do not report them here. Although an analytical compact expression for the error probability cannot be achieved, a numerical analysis is possible. We denote the information recovered using the quantum strategy as:
\begin{equation}
\mathcal{Q}:=1-H(p^{qua,PC}_{err}).
\end{equation}

\subsection*{Global classical limit}
A memory is constituted by an array of $D$ cells. In general information is stored as codewords constituted by the cells.  One can consider an information retrieval strategy consisting of simultaneous probing of the whole memory and a joint global POVM measurement at the output (see Fig. \ref{fig:scheme} B). If the input states are limited to tensor product of a state $\rho$ probing the single cell, the information per cell retrieved, $I^{D}$,  can be upper bounded\cite{Pirandola_2011a} using the Holevo Bound $\chi$:
\begin{equation}
I^{D}\leq \chi(\rho) := S(\rho)-\sum_i p_i S(\rho_i) \label{Hol}
\end{equation}
where $\rho=\sum_i p_i \rho_i$, $S(\rho)=-\text{Tr}(\rho\log{\rho})$ is the von-Neumann entropy\cite{Nielsen_2011}. In this case $i=0,1$, $\rho_i$ are the states defined in Eq.(\ref{channel}) and $p_i$ is the probability of the cell of being being prepared with either one of the values of $T_i$. In case of $D\to\infty$, that could be the case for very large memories, the  Holevo-Schumacher-Westmoreland (HSW) theorem \cite{Nielsen_2011} will assure that it exist a POVM such that the information will converge to $\chi$. 
Computing the bound in Eq. (\ref{Hol}) is a difficult task. The maximization becomes significantly easier if we restrict to the class of classical input states defined by Eq.(\ref{cla}). We can then define the maximum global classical information, $\chi_{cla}$, as the maximization of $\chi$ over all classical states:
\begin{equation}
\chi_{cla}:= \max_{\rho_{cla}} \chi(\rho) 
\end{equation}
Details on how the calculation of this limit is performed are reported in Appendix.

\section{Results}

\begin{figure*}
	\vspace{0.10cm}
	\includegraphics[width=2\columnwidth]{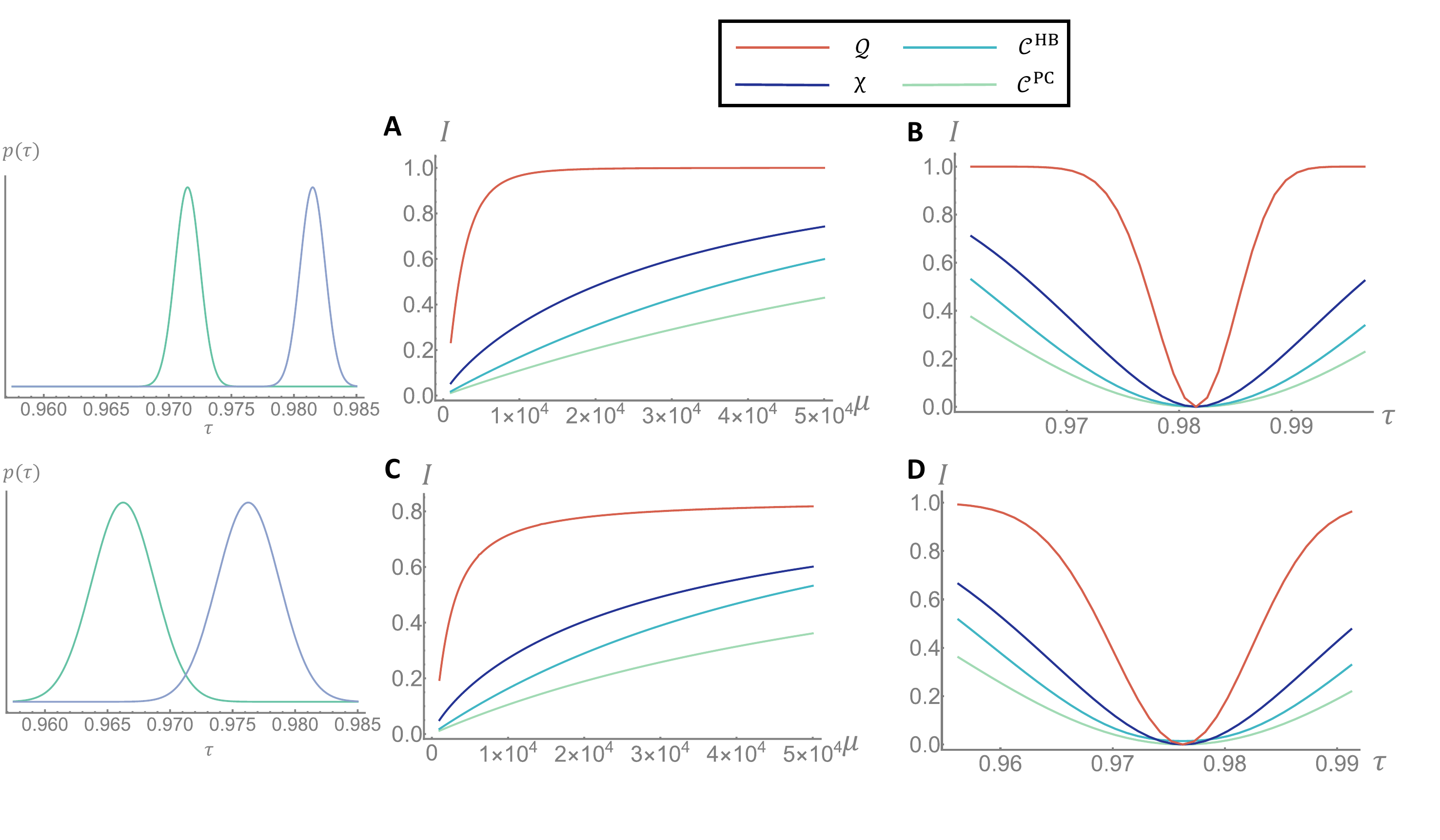}
	\caption{\label{fig:Fig2}\textbf{Comparison of information recovery.} We compare the information recovered, $I$ in bits, with the three classical strategies and the quantum one, described in the main text for two different configuration of the trasmittance distributions $g_{0}(\tau)$ and $g_{1}(\tau)$, both assumed gaussian. In the fist row, we fixed the mean value of $g_1$ to $\tau_1=0.982$ and the standard deviations of the distributions to $\sigma_1=\sigma_0=0.001$. In panel \textbf{A}, the mean value of $g_0$ is fixed to $\tau_0=0.972$ and the informations are showed as a function of the mean number of signal photons $\mu$. In all the panels we report the quantum recovered information $\mathcal{Q}$ in red, the global classical bound $\chi$ in dark blue, the local classical bound $\mathcal{C}^{HB}$ in light-blue and the photon counting classical performance  in light green. In panel \textbf{B} we fixed the photon number to $\mu=10^4$ and we showed the information in function of $\tau_0$. In the second row we showed the same dependency from $\mu$ in panel \textbf{C} for a different pair of transmittance distributions, namely we fixed $\tau_1=0.976$, $\tau_0=0.966$ and $\sigma_1=\sigma_0=0.0025$. In panel \textbf{D} we fixed $\mu=10^4$ and showed the dependency on $\tau_0$. }
\end{figure*}

\begin{figure*}
	\vspace{0.10cm}
	\includegraphics[width=2\columnwidth]{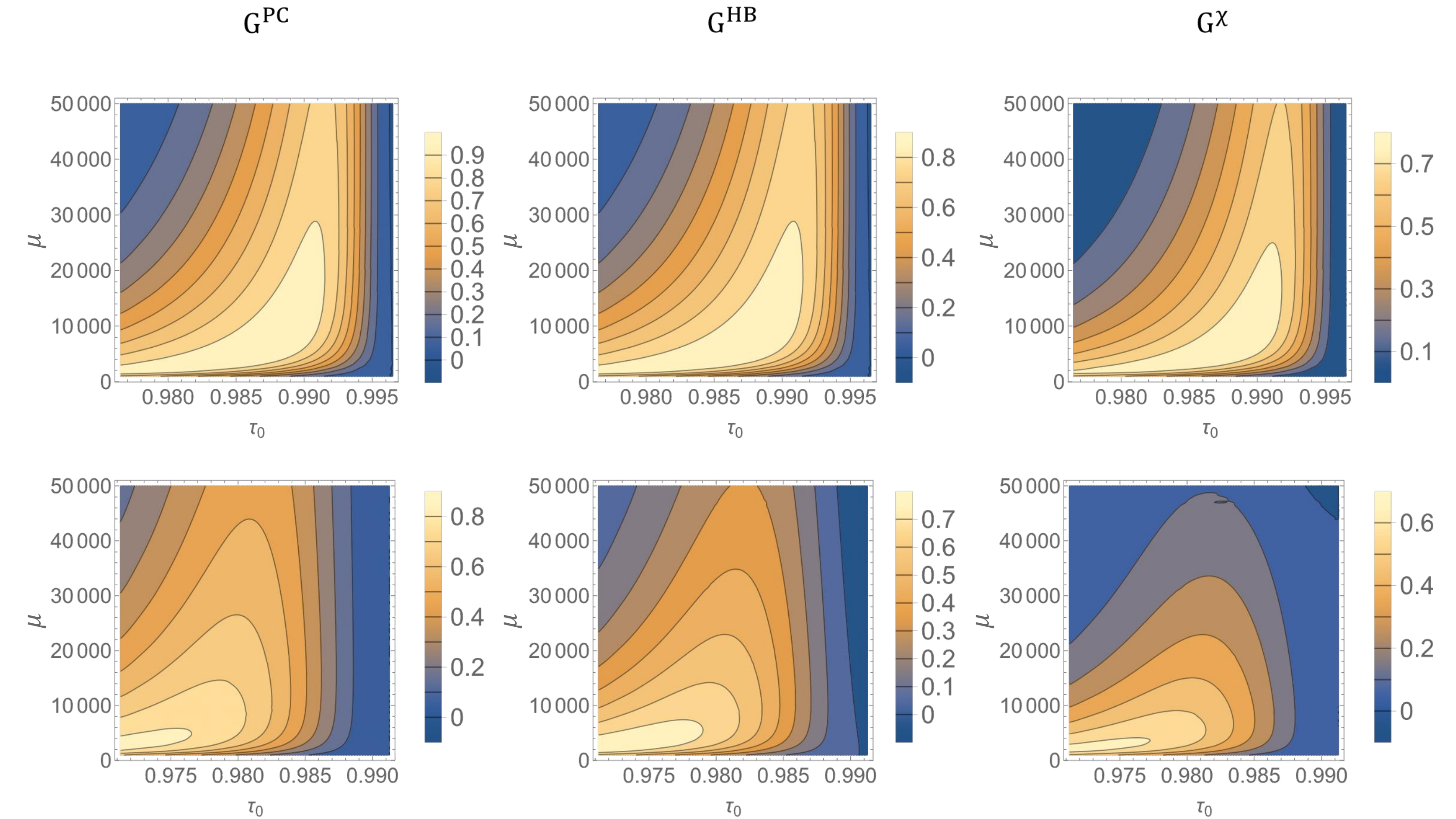}
	\caption{\label{fig:Fig3}\textbf{Quantum Gain.} We show the quantum gain $G$ in bits as a function of the mean number of signal photons $\mu$ and the mean value of the transmittance $\tau_0$ of one encoding distributions. In the upper row the other parameters are fixed to $\sigma_0=\sigma_1=0.001$ and $\tau_1=0.997$. Starting from the left: $G^{PC}$ is the quantum gain over to the performance of classical states an photon counting receiver, $G^{HB}$ over the optima bound of classical states considering local readout, and $G^{\chi}$ over the bound on the classical performance with global measurements. On the lower row we report the same figures of merit changing the parameters to $\sigma_0=\sigma_1=0.0025$ and $\tau_1=0.991$.}
\end{figure*}

\begin{figure}
	\vspace{0.10cm}
	\includegraphics[width=\columnwidth]{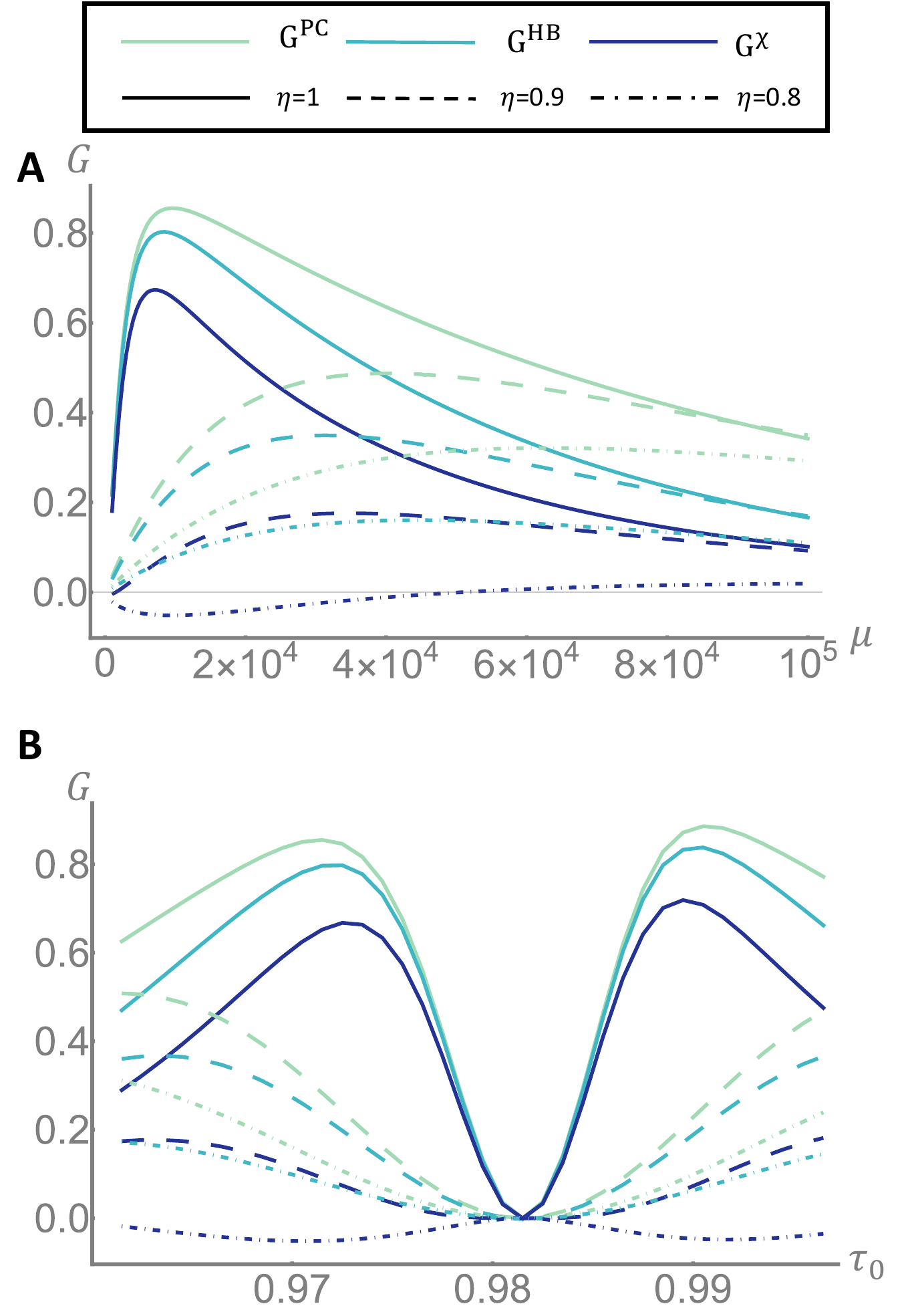}
	\caption{\label{fig:Fig4}\textbf{Inefficiency effect.} We report the quantum gain in bits over the three classical limit defined in the main text, for different values of overall efficiency $\eta$. In panel \textbf{A} we show the dependency of the gain on the main number of signal photons $\mu$. The case of perfect efficiency $\eta=1$ is reported in solid line, while the cases $\eta=0.9$ an $\eta=0.8$ are reported in dashed and dot dashed lines respectively. The distributions $g_0$ and $g_1$ are the same reported in the first row of Fig.(\ref{fig:Fig2}), namely $\tau_0=0.972$, $\tau_1=0.982$ and $\sigma_0=\sigma_1=0.001$.In panel \textbf{B} we fixed the number of photons to $\mu=10^4$ and we showed the dependency on the mean transmittance $\tau_0$.}
\end{figure}
To compare the quantum strategy with the three classical ones presented in the previous section we assume that the distributions $g_i(\tau)$ of the random variable $T_i$ are Gaussian and we denote the mean value as $\tau_i$ and the standard deviation as $\sigma_i$. In Fig.(\ref{fig:Fig2}) we report the results, in terms of bits of information recovered per cell, for two possible configurations of the transmittance's distributions. The first row shows a case in which the overlap between $g_0$ and $g_1$ is negligible. This means that, although the transmittance values are uncertain, in principle the value of the bit is codified in an unambiguous way, and a perfect measurement would be able to discriminate correctly its value. However, quantum fluctuations introduce further noise which reduce the actual distinguishability.
In panel \textbf{A}, we show the information recovered as a function of the mean number of signal photons $\mu$. The information recovered increases as the signal energy is increased, up until it saturates at the maximum amount of information for a single binary cell, i.e. 1 bit. In the range showed this saturation is only visible for the local quantum strategy, $\mathcal{Q}$ (reported in red), that reaches it earlier than any of the classical one, even the global capacity bound $\chi$. It shows that a the use of quantum resources allows a reliable recovery of the information with significantly less energy than otherwise needed. In Panel \textbf{B}, we fix the number of photons but let $\tau_0$ vary, while keeping fixed all the other parameter of the distributions. As expected, the recovered information is higher when $\tau_0$ is far from the fixed value of $\tau_1$ and starts to decrease when the two values get too close. However, in the quantum case, the strong degree of correlation of the source can be used to reduces the quantum fluctuations, which reflects in a much narrower low-informative region, the deep in Fig. \ref{fig:Fig2}\textbf{B}. 
The second row of Fig. \ref{fig:Fig2}, presents the case of a relevant overlapping of the initial distributions, as reported in the bottom-left box. In panel \textbf{C} and \textbf{D}, we show the dependence on the mean photon number $\mu$ and the transmittance $\tau_0$. Note that the information saturates at a value smaller than 1 bit (specifically 0.8) because the initial overlapping of the distributions. Of course, even a perfect measurement could not unravel the  initial ambiguous encoding. In panel \textbf{D} we see an effect similar to panel \textbf{B}, where the information recovered drops as $\tau_0$ approaches the fixed value of $\tau_1$. There is, however, a widening of the low-informative region w.r.t  the case of non-overlapping distributions. The quantum strategy present still a significant improvement in  this scenario were there is a greater part of indistinguishability not due to fluctuations.

We turn now our attention to the quantum gain defined as the difference:
\begin{equation}
G=\mathcal{Q}-\mathcal{C}
\end{equation}
where $\mathcal{C}$, can represent each one of the three classical bounds defined in Eq.s (\ref{CHB}), (\ref{CPC}) and (\ref{Hol}), in particular $G^{PC}$, $G^{HB}$ and $G^{\chi}$ is the quantum gain w.r.t $\mathcal{C}^{PC}$,  $\mathcal{C}^{HB}$ and $\chi_{cla}$ respectively. These quantities are reported in Fig. \ref{fig:Fig3} as a contour plot in function of the transmittance $\tau_0$ and the number of photons $\mu$, while the other parameters are fixed. In the first row of Fig. \ref{fig:Fig3}, the standard deviation of both distributions is fixed to $\sigma_0=\sigma_1=0.001$ which are small enough to reduce the initial overlapping. We see how the quantum gain in all three cases is relevant in most of the region analysed. The quantum strategy, based on photon counting measurement, performs very well against the same measurement strategy realized with classical probes, with the gain $G^{PC}$ reaching values higher then 0.9 bits. Thus, the use of a quantum probe allows recovering almost all the information in region where the same detection strategy with classical states would fail. A similar result is showed for the gain $G^{HB}$ over the bound on the the optimal local classical performance, although with slightly lower gains reaching a maximum of 0.8 bits. Even more remarkable, is the gain $G^{\chi}$ over the classical capacity, representing the bound on the information recovered per cell after a global measurement. The gain $G^{\chi}$ is significantly higher than zero in a wide region and reaches peaks between 0.7 and 0.8 bits. This shows how the improvement offered by quantum correlation cannot be substituted by classical encoding over large memories. In all three panels, the range of transmittance showing a significant advantage is wider for small number of photons, where quantum fluctuations are more relevant.
In the second row we show the effect of increasing the standard deviation to $\sigma_0=\sigma_1=0.0025$, leading to a large overlapping between the initial distributions in the region explored. We see in general a reduction of the gain due to the initial ambiguity of the encoding that limit the value of accessible information with to a value lower than 1 bit.

In experimental realizations, the main issue is the presence of optical losses from various sources. In the present scheme, the photon losses are accounted by the term $1-\eta$, with the efficiency $0\leq\eta\leq 1$. For the classical case the losses result simply in an effective reduction of the probing energy from $\mu$ to $\eta\mu$. In the quantum case, however, on top of the effective reduction of energy, losses have also an hindering effect on the correlations. The result of taking losses into account is then a reduction of the gain. In Fig. \ref{fig:Fig4} we report the the gain for the distribution case of non-overlapping initial distribution for different values of the efficiency. The panel \textbf{A} shows the gain as a function of the number of photons $\mu$. Although with a reduced gain, an advantage over the classical local bounds is preserved up to $\eta=0.8$ (20\% of losses), while the advantage is lost w.r.t. the classical capacity bound. It is worth noticing that the figure reported refers to gain per cell of information, so even a small fractions of information gained could result in a significant improvement over very large memories. In panel \textbf{B} the gain is reported as a function of the transmittance $\tau$. Beside the overall reduction of the gain, in presence of losses we observe a widening of the low-informative region in the $\tau_0$ range.
\begin{figure}
	\vspace{0.10cm}
	\includegraphics[width=\columnwidth]{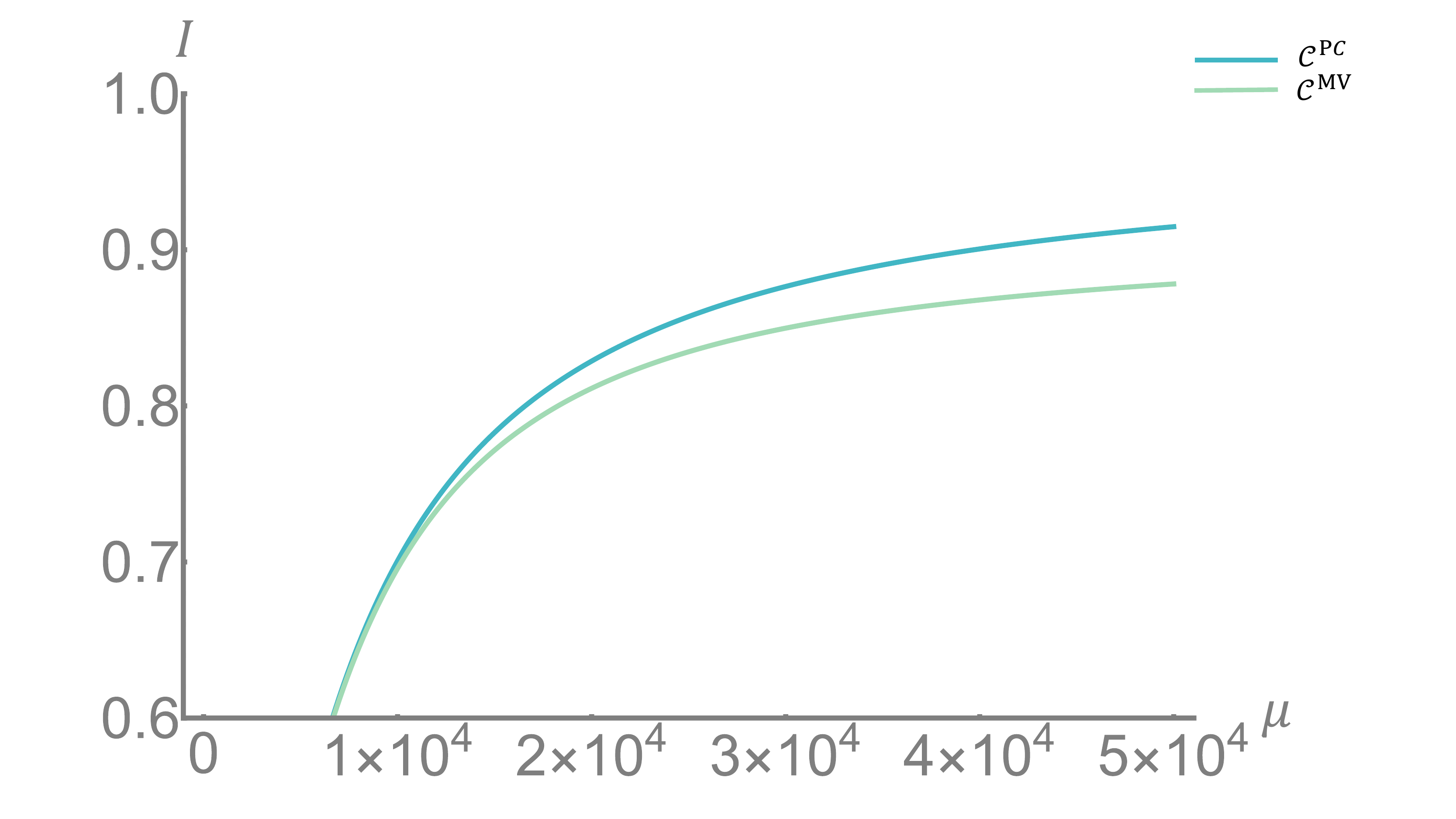}
	\caption{\label{fig:Fig5}\textbf{Post-processing comparison.} We show the bits of information retrieved by a classical transmitter and photon counting as a function of the mean number of signal photons $\mu$. We consider two maximum likelihood post processing, one $\mathcal{C}^{PC}$, with full information on the distributions of $T_0$, and $T_1$, as discussed in the main text, and the other $\mathcal{C}^{MV}$ using only the mean values of the distributions $\tau_0$ and $\tau_1$. The parameters are fixed to $\tau_0=925$,$\sigma_0=0.005$, $\tau_1=0.965$ and  $\sigma_1=0.01$. }
\end{figure}
Finally, in Fig. \ref{fig:Fig5} we compare the performance of the retrieval strategy proposed in this article with the one using only the two central values of transmittance $\tau_0$ and $\tau_1$, ignoring the distributions characterizing the memory. The information $\mathcal{C}^{PC}$ recovered using classical states and photon counting is compared with the information $\mathcal{C}^{MV}$ recovered with the same resources but using only the mean values of the distributions. As expected, the characterization of the memory, i.e. the knowledge of the distribution of the physical parameter used for the encoding, and its use in the decision algorithm brings an advantage in the readout.
\section{Discussion}

In this work we studied the effect of an imperfect characterization of a memory cell on its readout performance. In particular, we considered an optical memory storing a bit of information in two values of transmittance $T_0$ and $T_1$ that are not known with arbitrary precision, i.e. they are classical random variable with Gaussian distribution. In this scenario, we compared a specific quantum readout strategy, consisting of TMSV states transmitter and a photon counting receiver after the cell, with three classical informational limits:  The classical optimal bound for a single cell readout, the classical performance achievable with a photon counting receiver and the classical capacity limit taking into account collective measurement on a large memory.  Remarkably, the local quantum strategy reaches a notable advantage in terms of bit recovered per cell over all of them, even the last, global, one. Moreover, the advantage is retained for optical losses around $20\%$, which is particularly noticeable towards possible real applications, since losses represent the main limiting factor in many optical quantum sensing protocols.
Finally, we have shown that taking properly into account the parameter distributions in the decision algorithm, when available, allows to optimize in general the readout performance. 
While this work is mainly focused on a model of digital memory the results can be applied to different scenario involving convex superposition of loss channels. Some example are the conformance test \cite{Ortolano2021a} in the context of process monitoring, spectroscopy\cite{Shi_2020} and more in general any discrimination problem based on transmission/reflection optical measurement.

\vspace{6pt}

\section*{Funding} 
This work has been founded from the European Union’s Horizon 2020 Research and Innovation Action under Grant Agreement No. 862644 (Quantum readout techniques and technologies, QUARTET).
\vspace{6pt}
\appendix
\setcounter{equation}{0}
\renewcommand{\theequation}{A.\arabic{equation}}
\section{Appendix A}
\subsection{Classical Capacity}
As stated in the main text the imperfect memory readout problem can be seen as a problem of quantum channel discrimination. Given an input state  $\rho$, for the retrieval one has to discriminate between the two channels acting on $\rho$ as:
\begin{align}
\rho_0&=\mathbb{E}_{g_0}[(\mathcal{E}_\tau \otimes \mathcal{I}) \rho] \nonumber\\
\rho_1&=\mathbb{E}_{g_1}[(\mathcal{E}_\tau \otimes \mathcal{I}) \rho] \label{state}
\end{align}
In terms of information the single cell than can be seen as encoding the binary r.v. $X$ in the ensemble $\{p_i,\rho_i\}$, $i=0,1$. If a array of lenght $D$ of cells can be probed in parallel (see Fig.(\ref{fig:scheme}) in the main text), allowing joint measurement on all the output, a direct encoding, one bit per cell, may not be the more efficient storing strategy. In general the information will be encoded in codewords onto the array of cells. For the retrieval in the following we will consider strategies in which each cell is probed by a copy of a state $\rho$, so that the total probing state is of the form $\rho^{\otimes  D}$.  The multicell storage/retrieval of information is a transfer of information  through the memory, characterized in terms of quantum channels, and the maximum rate of information $I$ that can be be retrieved per single channel use (cell of the memory) is limited by the Holevo Bound:
 \begin{equation}
 I\leq\chi(\rho):=S(\rho_s)-\sum_i p_i S(\rho_i) \label{holevo}
 \end{equation}
 where $\rho_s=\sum_i p_i\rho_i$ and $S(\rho)=-Tr(\rho \log \rho)$ is the von-Neumann entropy \cite{Nielsen_2011}. The less or equal sign in equation reflects the fact that for finite array of lenght $D$ the convergence is not guaranteed. On the other hand for an infinite number of cells the the HWS theorem guarantees the convergence of the information recovered by the optimal strategy to the Holevo quantity \cite{Nielsen_2011}. The information recovered depends on the input state $\rho$. As already done in the main text we will impose a constraint on the input states allowed in the form of a fixed signal number of photons $\mu$. The optimization of the quantity would give an ultimate limit on the rate of information per bit that can be stored and retrieved accurately. However this optimization is not easy to perform on the full space of states. The problem can be solved more easily if the optimization is restricted to the class of classical states defined in Eq.(\ref{cla}). Solving this optimization will yield the \textit{classical capacity} of the imperfect memory, $\chi^{cla}$: 
\begin{equation}
\chi^{cla}:=\max_{\rho_{cla}} \chi(\rho_{cla}) 
\end{equation} 
 To find $\chi^{cla}$ we will first find the Holevo quantity for a single mode coherent state and then, following a similar proof given in Ref.\cite{Pirandola_2011a} for a perfect memory, we will show how the quantity found is equal to the classical capacity.
 
\subsection{Single mode coherent state}
Consider a single mode coherent state $|\alpha\rangle\langle\alpha|$, without idler modes, as an input for the readout procedure. In this case the constraint on the number of photons gives the condition $|\alpha|^2=\mu$. To compute the Holevo quantity for this state, $\chi^{\alpha}$, we first use the fact that pure loss channels  map coherent states into coherent states, $\mathcal{E}_\tau(|\alpha\rangle\langle\alpha|)=|\sqrt{\tau}\alpha\rangle\langle\sqrt{\tau}\alpha|$, to compute the output states in Eq.(\ref{state}):
\begin{align}
\rho^\alpha_0&=\mathbb{E}_{g_0}[|\sqrt{\tau}\alpha\rangle\langle\sqrt{\tau}\alpha|]=\int d\tau g_0(\tau) |\sqrt{\tau}\alpha\rangle\langle\sqrt{\tau}\alpha| \nonumber\\
\rho^\alpha_1&=\mathbb{E}_{g_1}[|\sqrt{\tau}\alpha\rangle\langle\sqrt{\tau}\alpha|]=\int d\tau g_1(\tau) |\sqrt{\tau}\alpha\rangle\langle\sqrt{\tau}\alpha| \label{statecoh}
\end{align}
where, since there are no idler modes, we omitted them from the notation. We have then:
\begin{equation}
\chi^{\alpha}=S(p_0\rho_0^\alpha+p_1\rho_1^\alpha)-\sum_i p_i S(\rho_i^\alpha)\label{chialpha}
\end{equation}
The computation of the entropy is complicated by the fact that the distribution $g_i$ are in general continuous ones. To overcome this problem, we performed a uniform discretization of the distribution to some appropriate dimension $k$, so that the integrals over $\tau$ in Eq.(\ref{statecoh}) are substituted by finite sums, $\int d\tau \to \sum_i^k$. Under this approximation is easy to see how each entropy term in Eq.(\ref{chialpha}) can be rewritten as some convex combination of a finite subset of maximum dimension $2k$ of coherent states. The calculation of $\chi^{\alpha}$ than can be reduced to the calculations of terms in the form:
\begin{equation}
S\Bigg(\sum_{i=1}^K q_i |\sqrt{\tau_i}\alpha\rangle\langle\sqrt{\tau_i}\alpha|\Bigg)
\end{equation}

where $q_i$ are suitable probability coefficients depending on the initial distributions $g_i$ and the term considered. For the first term on the right hand side of Eq.(\ref{chialpha}) the coefficients $q_i$ will depend also on the ensemble probabilities $p_i$ and the sum will in general be on $K=2k$ terms. For the two entropy contributions on the summation the sum will go over $K=k$ terms. 

In general, the set $\{|\sqrt{\tau_i}\alpha\rangle\}$ is a non-orthogonal basis of a $K$-dimensional Hilbert space. For any state $\rho=\sum_{i=1}^K q_i |\sqrt{\tau_i}\alpha\rangle\langle\sqrt{\tau_i}\alpha|$, we have:
\begin{equation}
S(\rho)=S(QG) \label{or}
\end{equation}
where $Q=\text{Diag}[q_i]$ and  $G(i,j)=G(j,i)=\langle\sqrt{\tau_i}\alpha|\sqrt{\tau_j}\alpha\rangle$ are the elements of the Gram matrix $G$. In fact, we can write \cite{Gagatsos_2016}:
\begin{equation}
S(\rho)=-\frac{\partial}{\partial n}\text{Tr}(\rho^n)  \Big|_{n=1} \label{rm}
\end{equation}
And
\begin{widetext}
\begin{align}
&\text{Tr}(\rho^n)=\sum_{i_1,...,i_n=1}^K  q_{i_1} \cdots q_{i_n} \langle\sqrt{\tau_{i_1}}\alpha|\sqrt{\tau_{i_1}}\alpha\rangle \cdots \langle\sqrt{\tau_{i_n}}\alpha|\sqrt{\tau_{i_1}}\alpha\rangle = \nonumber \\
&=\sum_{i_2, ...,i_n=1}^K q_{i_2} \cdots q_{i_n} \langle\sqrt{\tau_{i_2}}\alpha|\sqrt{\tau_{i_3}}\alpha\rangle \cdots \langle\sqrt{\tau_{i_{n-1}}}\alpha|\sqrt{\tau_{i_n}}\alpha\rangle GQG(i_n,i_2) = \nonumber \\
&=\sum_{i_n=1}^K q_{i_n} G(QG)^{n-1}(i_n,i_n)=\text{Tr}[(QG)^n] \label{tr}
\end{align}
\end{widetext}
Combining the results of Eq.(\ref{rm}) and Eq.(\ref{tr}), we get the equality in Eq.(\ref{or}). 

While a closed form may not be available for a generic probability distribution and for arbitrary $k$, Eq.(\ref{or}) allows to skip the orthogonalization of the subspace, which would represent a computational heavy task for large values of $k$. A numerical evaluation of $\chi^\alpha$ can be performed using Eq.(\ref{or}-\ref{rm}). The Holevo quantity $\chi^\alpha$ for the continuous set can be recovered as the limit for $k\to\infty$ of $\chi^{cla}(k)$. In the following section we will show how $\chi^\alpha$ coincides with the classical capacity $\chi^{cla}$.

\subsection{Saturation of capacity by single mode coherent state} \label{Saturation of capacity by single mode coherent state} 

To prove that the classical capacity can be computed using a single mode coherent transmitter we can use the argument used in Ref.\cite{Pirandola_2011a} that we report in the following. 

Consider the class $\mathcal{P}$ of pure coherent transmitters. The class of mixed states formed by positive superposition of elements of $\mathcal{P}$ constitutes the class of classical states. Using the convexity on $\rho$ of the Holevo information $\chi$, it can be proved, similarly to what was done in \cite{Pirandola_2011a}, that, if the capacity of the class $\mathcal{P}$, $\chi^P$, is concave in the number of photons $\mu$, it must be larger or equal to the capacity of whole class of classical states:
\begin{equation}
\chi^P\geq\chi^{cla} \label{chipure}
\end{equation}
Given that $\mathcal{P}$ belongs is a subclass of the classical states, the capacity of $\mathcal{P}$ cannot be larger than the one of classical states, so that Eq.(\ref{chipure}) is an equality. The capacity of classical states is then saturated by pure coherent states. The concavity of $\chi_{\mathcal{P}}$ can be checked numerically.

To complete the proof, one must simply show that a single coherent mode saturates the capacity of $\mathcal{P}$. This can be done\cite{Pirandola_2011a} using the fact that acting with unitary transformations, that don't change the von Neumann entropy, on the output states of any pure classical input one can rearrange the signal photons in a single mode obtaining the same result as a single mode input.
\bibliography{bib}

\begin{thebibliography}{49}%
\makeatletter
\providecommand \@ifxundefined [1]{%
 \@ifx{#1\undefined}
}%
\providecommand \@ifnum [1]{%
 \ifnum #1\expandafter \@firstoftwo
 \else \expandafter \@secondoftwo
 \fi
}%
\providecommand \@ifx [1]{%
 \ifx #1\expandafter \@firstoftwo
 \else \expandafter \@secondoftwo
 \fi
}%
\providecommand \natexlab [1]{#1}%
\providecommand \enquote  [1]{``#1''}%
\providecommand \bibnamefont  [1]{#1}%
\providecommand \bibfnamefont [1]{#1}%
\providecommand \citenamefont [1]{#1}%
\providecommand \href@noop [0]{\@secondoftwo}%
\providecommand \href [0]{\begingroup \@sanitize@url \@href}%
\providecommand \@href[1]{\@@startlink{#1}\@@href}%
\providecommand \@@href[1]{\endgroup#1\@@endlink}%
\providecommand \@sanitize@url [0]{\catcode `\\12\catcode `\$12\catcode
  `\&12\catcode `\#12\catcode `\^12\catcode `\_12\catcode `\%12\relax}%
\providecommand \@@startlink[1]{}%
\providecommand \@@endlink[0]{}%
\providecommand \url  [0]{\begingroup\@sanitize@url \@url }%
\providecommand \@url [1]{\endgroup\@href {#1}{\urlprefix }}%
\providecommand \urlprefix  [0]{URL }%
\providecommand \Eprint [0]{\href }%
\providecommand \doibase [0]{https://doi.org/}%
\providecommand \selectlanguage [0]{\@gobble}%
\providecommand \bibinfo  [0]{\@secondoftwo}%
\providecommand \bibfield  [0]{\@secondoftwo}%
\providecommand \translation [1]{[#1]}%
\providecommand \BibitemOpen [0]{}%
\providecommand \bibitemStop [0]{}%
\providecommand \bibitemNoStop [0]{.\EOS\space}%
\providecommand \EOS [0]{\spacefactor3000\relax}%
\providecommand \BibitemShut  [1]{\csname bibitem#1\endcsname}%
\let\auto@bib@innerbib\@empty
\bibitem [{\citenamefont {Giovannetti}\ \emph {et~al.}(2011)\citenamefont
  {Giovannetti}, \citenamefont {Lloyd},\ and\ \citenamefont
  {Maccone}}]{Giovannetti_2011}%
  \BibitemOpen
  \bibfield  {author} {\bibinfo {author} {\bibfnamefont {V.}~\bibnamefont
  {Giovannetti}}, \bibinfo {author} {\bibfnamefont {S.}~\bibnamefont {Lloyd}},\
  and\ \bibinfo {author} {\bibfnamefont {L.}~\bibnamefont {Maccone}},\
  }\bibfield  {title} {\bibinfo {title} {Advances in quantum metrology},\
  }\href {https://doi.org/10.1038/nphoton.2011.35} {\bibfield  {journal}
  {\bibinfo  {journal} {Nature Photonics}\ }\textbf {\bibinfo {volume} {5}},\
  \bibinfo {pages} {222 EP } (\bibinfo {year} {2011})},\ \bibinfo {note}
  {review Article}\BibitemShut {NoStop}%
\bibitem [{\citenamefont {Pirandola}\ \emph {et~al.}(2018)\citenamefont
  {Pirandola}, \citenamefont {Bardhan}, \citenamefont {Gehring}, \citenamefont
  {Weedbrook},\ and\ \citenamefont {Lloyd}}]{Pirandola_2018}%
  \BibitemOpen
  \bibfield  {author} {\bibinfo {author} {\bibfnamefont {S.}~\bibnamefont
  {Pirandola}}, \bibinfo {author} {\bibfnamefont {B.~R.}\ \bibnamefont
  {Bardhan}}, \bibinfo {author} {\bibfnamefont {T.}~\bibnamefont {Gehring}},
  \bibinfo {author} {\bibfnamefont {C.}~\bibnamefont {Weedbrook}},\ and\
  \bibinfo {author} {\bibfnamefont {S.}~\bibnamefont {Lloyd}},\ }\bibfield
  {title} {\bibinfo {title} {Advances in photonic quantum sensing},\ }\href
  {https://doi.org/10.1038/s41566-018-0301-6} {\bibfield  {journal} {\bibinfo
  {journal} {Nature Photonics}\ }\textbf {\bibinfo {volume} {12}},\ \bibinfo
  {pages} {724} (\bibinfo {year} {2018})}\BibitemShut {NoStop}%
\bibitem [{\citenamefont {Berchera}\ and\ \citenamefont
  {Degiovanni}(2019)}]{Berchera_2019}%
  \BibitemOpen
  \bibfield  {author} {\bibinfo {author} {\bibfnamefont {I.~R.}\ \bibnamefont
  {Berchera}}\ and\ \bibinfo {author} {\bibfnamefont {I.~P.}\ \bibnamefont
  {Degiovanni}},\ }\bibfield  {title} {\bibinfo {title} {Quantum imaging with
  sub-poissonian light: challenges and perspectives in optical metrology},\
  }\href {https://doi.org/10.1088/1681-7575/aaf7b2} {\bibfield  {journal}
  {\bibinfo  {journal} {Metrologia}\ }\textbf {\bibinfo {volume} {56}},\
  \bibinfo {pages} {024001} (\bibinfo {year} {2019})}\BibitemShut {NoStop}%
\bibitem [{\citenamefont {Giovannetti}\ \emph {et~al.}(2004)\citenamefont
  {Giovannetti}, \citenamefont {Lloyd},\ and\ \citenamefont
  {Maccone}}]{Giovannetti_2004}%
  \BibitemOpen
  \bibfield  {author} {\bibinfo {author} {\bibfnamefont {V.}~\bibnamefont
  {Giovannetti}}, \bibinfo {author} {\bibfnamefont {S.}~\bibnamefont {Lloyd}},\
  and\ \bibinfo {author} {\bibfnamefont {L.}~\bibnamefont {Maccone}},\
  }\bibfield  {title} {\bibinfo {title} {Quantum-enhanced measurements: Beating
  the standard quantum limit},\ }\href
  {https://doi.org/10.1126/science.1104149} {\bibfield  {journal} {\bibinfo
  {journal} {Science}\ }\textbf {\bibinfo {volume} {306}},\ \bibinfo {pages}
  {1330} (\bibinfo {year} {2004})},\ \Eprint
  {https://arxiv.org/abs/https://science.sciencemag.org/content/306/5700/1330.full.pdf}
  {https://science.sciencemag.org/content/306/5700/1330.full.pdf} \BibitemShut
  {NoStop}%
\bibitem [{\citenamefont {Schnabel}(2017)}]{Schnabel_2017}%
  \BibitemOpen
  \bibfield  {author} {\bibinfo {author} {\bibfnamefont {R.}~\bibnamefont
  {Schnabel}},\ }\bibfield  {title} {\bibinfo {title} {Squeezed states of light
  and their applications in laser interferometers},\ }\href
  {https://doi.org/https://doi.org/10.1016/j.physrep.2017.04.001} {\bibfield
  {journal} {\bibinfo  {journal} {Physics Reports}\ }\textbf {\bibinfo {volume}
  {684}},\ \bibinfo {pages} {1 } (\bibinfo {year} {2017})},\ \bibinfo {note}
  {squeezed states of light and their applications in laser
  interferometers}\BibitemShut {NoStop}%
\bibitem [{\citenamefont {Sch\"{a}fermeier}\ \emph {et~al.}(2018)\citenamefont
  {Sch\"{a}fermeier}, \citenamefont {Je\v{z}ek}, \citenamefont {Madsen},
  \citenamefont {Gehring},\ and\ \citenamefont {Andersen}}]{Schafermeier_2018}%
  \BibitemOpen
  \bibfield  {author} {\bibinfo {author} {\bibfnamefont {C.}~\bibnamefont
  {Sch\"{a}fermeier}}, \bibinfo {author} {\bibfnamefont {M.}~\bibnamefont
  {Je\v{z}ek}}, \bibinfo {author} {\bibfnamefont {L.~S.}\ \bibnamefont
  {Madsen}}, \bibinfo {author} {\bibfnamefont {T.}~\bibnamefont {Gehring}},\
  and\ \bibinfo {author} {\bibfnamefont {U.~L.}\ \bibnamefont {Andersen}},\
  }\bibfield  {title} {\bibinfo {title} {Deterministic phase measurements
  exhibiting super-sensitivity and super-resolution},\ }\href
  {https://doi.org/10.1364/OPTICA.5.000060} {\bibfield  {journal} {\bibinfo
  {journal} {Optica}\ }\textbf {\bibinfo {volume} {5}},\ \bibinfo {pages} {60}
  (\bibinfo {year} {2018})}\BibitemShut {NoStop}%
\bibitem [{\citenamefont {Ortolano}\ \emph {et~al.}(2019)\citenamefont
  {Ortolano}, \citenamefont {Ruo-Berchera},\ and\ \citenamefont
  {Predazzi}}]{Ortolano_2019}%
  \BibitemOpen
  \bibfield  {author} {\bibinfo {author} {\bibfnamefont {G.}~\bibnamefont
  {Ortolano}}, \bibinfo {author} {\bibfnamefont {I.}~\bibnamefont
  {Ruo-Berchera}},\ and\ \bibinfo {author} {\bibfnamefont {E.}~\bibnamefont
  {Predazzi}},\ }\bibfield  {title} {\bibinfo {title} {Quantum enhanced imaging
  of nonuniform refractive profiles},\ }\href@noop {} {\bibfield  {journal}
  {\bibinfo  {journal} {International Journal of Quantum Information}\ }\textbf
  {\bibinfo {volume} {17}},\ \bibinfo {pages} {1941010} (\bibinfo {year}
  {2019})}\BibitemShut {NoStop}%
\bibitem [{\citenamefont {Monras}\ and\ \citenamefont
  {Paris}(2007)}]{Monras_2007}%
  \BibitemOpen
  \bibfield  {author} {\bibinfo {author} {\bibfnamefont {A.}~\bibnamefont
  {Monras}}\ and\ \bibinfo {author} {\bibfnamefont {M.~G.~A.}\ \bibnamefont
  {Paris}},\ }\bibfield  {title} {\bibinfo {title} {Optimal quantum estimation
  of loss in bosonic channels},\ }\href
  {https://doi.org/10.1103/PhysRevLett.98.160401} {\bibfield  {journal}
  {\bibinfo  {journal} {Physical Review Letter}\ }\textbf {\bibinfo {volume}
  {98}},\ \bibinfo {pages} {160401} (\bibinfo {year} {2007})}\BibitemShut
  {NoStop}%
\bibitem [{\citenamefont {Adesso}\ \emph {et~al.}(2009)\citenamefont {Adesso},
  \citenamefont {Dell'Anno}, \citenamefont {De~Siena}, \citenamefont
  {Illuminati},\ and\ \citenamefont {Souza}}]{Adesso_2009}%
  \BibitemOpen
  \bibfield  {author} {\bibinfo {author} {\bibfnamefont {G.}~\bibnamefont
  {Adesso}}, \bibinfo {author} {\bibfnamefont {F.}~\bibnamefont {Dell'Anno}},
  \bibinfo {author} {\bibfnamefont {S.}~\bibnamefont {De~Siena}}, \bibinfo
  {author} {\bibfnamefont {F.}~\bibnamefont {Illuminati}},\ and\ \bibinfo
  {author} {\bibfnamefont {L.~A.~M.}\ \bibnamefont {Souza}},\ }\bibfield
  {title} {\bibinfo {title} {Optimal estimation of losses at the ultimate
  quantum limit with non-gaussian states},\ }\href
  {https://doi.org/10.1103/PhysRevA.79.040305} {\bibfield  {journal} {\bibinfo
  {journal} {Physical Review A}\ }\textbf {\bibinfo {volume} {79}},\ \bibinfo
  {pages} {040305} (\bibinfo {year} {2009})}\BibitemShut {NoStop}%
\bibitem [{\citenamefont {Losero}\ \emph {et~al.}(2018)\citenamefont {Losero},
  \citenamefont {Ruo-Berchera}, \citenamefont {Meda}, \citenamefont {Avella},\
  and\ \citenamefont {Genovese}}]{Losero_2018}%
  \BibitemOpen
  \bibfield  {author} {\bibinfo {author} {\bibfnamefont {E.}~\bibnamefont
  {Losero}}, \bibinfo {author} {\bibfnamefont {I.}~\bibnamefont
  {Ruo-Berchera}}, \bibinfo {author} {\bibfnamefont {A.}~\bibnamefont {Meda}},
  \bibinfo {author} {\bibfnamefont {A.}~\bibnamefont {Avella}},\ and\ \bibinfo
  {author} {\bibfnamefont {M.}~\bibnamefont {Genovese}},\ }\bibfield  {title}
  {\bibinfo {title} {Unbiased estimation of an optical loss at the ultimate
  quantum limit with twin-beams},\ }\href
  {https://doi.org/10.1038/s41598-018-25501-w} {\bibfield  {journal} {\bibinfo
  {journal} {Scientific Reports}\ }\textbf {\bibinfo {volume} {8}},\ \bibinfo
  {pages} {7431} (\bibinfo {year} {2018})}\BibitemShut {NoStop}%
\bibitem [{\citenamefont {{Brida}}\ \emph {et~al.}(2010)\citenamefont
  {{Brida}}, \citenamefont {{Genovese}},\ and\ \citenamefont {{Ruo
  Berchera}}}]{Brida_2010}%
  \BibitemOpen
  \bibfield  {author} {\bibinfo {author} {\bibfnamefont {G.}~\bibnamefont
  {{Brida}}}, \bibinfo {author} {\bibfnamefont {M.}~\bibnamefont
  {{Genovese}}},\ and\ \bibinfo {author} {\bibfnamefont {I.}~\bibnamefont {{Ruo
  Berchera}}},\ }\bibfield  {title} {\bibinfo {title} {{Experimental
  realization of sub-shot-noise quantum imaging}},\ }\href@noop {} {\bibfield
  {journal} {\bibinfo  {journal} {Nature Photonics}\ }\textbf {\bibinfo
  {volume} {4}},\ \bibinfo {pages} {227} (\bibinfo {year} {2010})}\BibitemShut
  {NoStop}%
\bibitem [{\citenamefont {Genovese}(2016)}]{Genovese_2016}%
  \BibitemOpen
  \bibfield  {author} {\bibinfo {author} {\bibfnamefont {M.}~\bibnamefont
  {Genovese}},\ }\bibfield  {title} {\bibinfo {title} {Real applications of
  quantum imaging},\ }\href@noop {} {\bibfield  {journal} {\bibinfo  {journal}
  {Journal of Optics}\ }\textbf {\bibinfo {volume} {18}},\ \bibinfo {pages}
  {073002} (\bibinfo {year} {2016})}\BibitemShut {NoStop}%
\bibitem [{\citenamefont {Aasi}\ \emph {et~al.}(2013)\citenamefont {Aasi} \emph
  {et~al.}}]{Aasi_2013}%
  \BibitemOpen
  \bibfield  {author} {\bibinfo {author} {\bibfnamefont {J.}~\bibnamefont
  {Aasi}} \emph {et~al.},\ }\bibfield  {title} {\bibinfo {title} {Enhanced
  sensitivity of the ligo gravitational wave detector by using squeezed states
  of light},\ }\href {https://doi.org/10.1038/nphoton.2013.177} {\bibfield
  {journal} {\bibinfo  {journal} {Nature Photonics}\ }\textbf {\bibinfo
  {volume} {7}},\ \bibinfo {pages} {613 EP } (\bibinfo {year}
  {2013})}\BibitemShut {NoStop}%
\bibitem [{\citenamefont {Ruo~Berchera}\ \emph {et~al.}(2013)\citenamefont
  {Ruo~Berchera}, \citenamefont {Degiovanni}, \citenamefont {Olivares},\ and\
  \citenamefont {Genovese}}]{Berchera_2013}%
  \BibitemOpen
  \bibfield  {author} {\bibinfo {author} {\bibfnamefont {I.}~\bibnamefont
  {Ruo~Berchera}}, \bibinfo {author} {\bibfnamefont {I.~P.}\ \bibnamefont
  {Degiovanni}}, \bibinfo {author} {\bibfnamefont {S.}~\bibnamefont
  {Olivares}},\ and\ \bibinfo {author} {\bibfnamefont {M.}~\bibnamefont
  {Genovese}},\ }\bibfield  {title} {\bibinfo {title} {Quantum light in coupled
  interferometers for quantum gravity tests},\ }\href
  {https://doi.org/10.1103/PhysRevLett.110.213601} {\bibfield  {journal}
  {\bibinfo  {journal} {Physical Review Letter}\ }\textbf {\bibinfo {volume}
  {110}},\ \bibinfo {pages} {213601} (\bibinfo {year} {2013})}\BibitemShut
  {NoStop}%
\bibitem [{\citenamefont {Pradyumna}\ \emph {et~al.}(2020)\citenamefont
  {Pradyumna}, \citenamefont {Losero}, \citenamefont {Ruo-Berchera},
  \citenamefont {Traina}, \citenamefont {Zucco}, \citenamefont {Jacobsen},
  \citenamefont {Andersen}, \citenamefont {Degiovanni}, \citenamefont
  {Genovese},\ and\ \citenamefont {Gehring}}]{Pradyumna_2020}%
  \BibitemOpen
  \bibfield  {author} {\bibinfo {author} {\bibfnamefont {S.~T.}\ \bibnamefont
  {Pradyumna}}, \bibinfo {author} {\bibfnamefont {E.}~\bibnamefont {Losero}},
  \bibinfo {author} {\bibfnamefont {I.}~\bibnamefont {Ruo-Berchera}}, \bibinfo
  {author} {\bibfnamefont {P.}~\bibnamefont {Traina}}, \bibinfo {author}
  {\bibfnamefont {M.}~\bibnamefont {Zucco}}, \bibinfo {author} {\bibfnamefont
  {C.~S.}\ \bibnamefont {Jacobsen}}, \bibinfo {author} {\bibfnamefont {U.~L.}\
  \bibnamefont {Andersen}}, \bibinfo {author} {\bibfnamefont {I.~P.}\
  \bibnamefont {Degiovanni}}, \bibinfo {author} {\bibfnamefont
  {M.}~\bibnamefont {Genovese}},\ and\ \bibinfo {author} {\bibfnamefont
  {T.}~\bibnamefont {Gehring}},\ }\bibfield  {title} {\bibinfo {title} {Twin
  beam quantum-enhanced correlated interferometry for testing fundamental
  physics},\ }\href {https://doi.org/10.1038/s42005-020-0368-5} {\bibfield
  {journal} {\bibinfo  {journal} {Communications Physics}\ }\textbf {\bibinfo
  {volume} {3}},\ \bibinfo {pages} {104} (\bibinfo {year} {2020})}\BibitemShut
  {NoStop}%
\bibitem [{\citenamefont {Helstrom}(1976)}]{Helstrom_1976}%
  \BibitemOpen
  \bibfield  {author} {\bibinfo {author} {\bibfnamefont {C.}~\bibnamefont
  {Helstrom}},\ }\href@noop {} {\emph {\bibinfo {title} {Quantum detection and
  estimation theory}}}\ (\bibinfo  {publisher} {Academic Press},\ \bibinfo
  {address} {New York},\ \bibinfo {year} {1976})\BibitemShut {NoStop}%
\bibitem [{\citenamefont {Chefles}\ and\ \citenamefont
  {Barnett}(1998)}]{Chefles_1998}%
  \BibitemOpen
  \bibfield  {author} {\bibinfo {author} {\bibfnamefont {A.}~\bibnamefont
  {Chefles}}\ and\ \bibinfo {author} {\bibfnamefont {S.~M.}\ \bibnamefont
  {Barnett}},\ }\bibfield  {title} {\bibinfo {title} {Quantum state separation,
  unambiguous discrimination and exact cloning},\ }\href
  {https://doi.org/10.1088/0305-4470/31/50/007} {\bibfield  {journal} {\bibinfo
   {journal} {Journal of Physics A: Mathematical and General}\ }\textbf
  {\bibinfo {volume} {31}},\ \bibinfo {pages} {10097} (\bibinfo {year}
  {1998})}\BibitemShut {NoStop}%
\bibitem [{\citenamefont {Lloyd}(2008)}]{Lloyd_2008}%
  \BibitemOpen
  \bibfield  {author} {\bibinfo {author} {\bibfnamefont {S.}~\bibnamefont
  {Lloyd}},\ }\bibfield  {title} {\bibinfo {title} {Enhanced sensitivity of
  photodetection via quantum illumination},\ }\href
  {https://doi.org/10.1126/science.1160627} {\bibfield  {journal} {\bibinfo
  {journal} {Science}\ }\textbf {\bibinfo {volume} {321}},\ \bibinfo {pages}
  {1463} (\bibinfo {year} {2008})}\BibitemShut {NoStop}%
\bibitem [{\citenamefont {Tan}\ \emph {et~al.}(2008)\citenamefont {Tan},
  \citenamefont {Erkmen}, \citenamefont {Giovannetti}, \citenamefont {Guha},
  \citenamefont {Lloyd}, \citenamefont {Maccone}, \citenamefont {Pirandola},\
  and\ \citenamefont {Shapiro}}]{Tan_2008}%
  \BibitemOpen
  \bibfield  {author} {\bibinfo {author} {\bibfnamefont {S.-H.}\ \bibnamefont
  {Tan}}, \bibinfo {author} {\bibfnamefont {B.~I.}\ \bibnamefont {Erkmen}},
  \bibinfo {author} {\bibfnamefont {V.}~\bibnamefont {Giovannetti}}, \bibinfo
  {author} {\bibfnamefont {S.}~\bibnamefont {Guha}}, \bibinfo {author}
  {\bibfnamefont {S.}~\bibnamefont {Lloyd}}, \bibinfo {author} {\bibfnamefont
  {L.}~\bibnamefont {Maccone}}, \bibinfo {author} {\bibfnamefont
  {S.}~\bibnamefont {Pirandola}},\ and\ \bibinfo {author} {\bibfnamefont
  {J.~H.}\ \bibnamefont {Shapiro}},\ }\bibfield  {title} {\bibinfo {title}
  {Quantum illumination with gaussian states},\ }\href
  {https://doi.org/10.1103/PhysRevLett.101.253601} {\bibfield  {journal}
  {\bibinfo  {journal} {Physical Review Letter}\ }\textbf {\bibinfo {volume}
  {101}},\ \bibinfo {pages} {253601} (\bibinfo {year} {2008})}\BibitemShut
  {NoStop}%
\bibitem [{\citenamefont {Lopaeva}\ \emph {et~al.}(2013)\citenamefont
  {Lopaeva}, \citenamefont {Ruo~Berchera}, \citenamefont {Degiovanni},
  \citenamefont {Olivares}, \citenamefont {Brida},\ and\ \citenamefont
  {Genovese}}]{Lopaeva_2013}%
  \BibitemOpen
  \bibfield  {author} {\bibinfo {author} {\bibfnamefont {E.~D.}\ \bibnamefont
  {Lopaeva}}, \bibinfo {author} {\bibfnamefont {I.}~\bibnamefont
  {Ruo~Berchera}}, \bibinfo {author} {\bibfnamefont {I.~P.}\ \bibnamefont
  {Degiovanni}}, \bibinfo {author} {\bibfnamefont {S.}~\bibnamefont
  {Olivares}}, \bibinfo {author} {\bibfnamefont {G.}~\bibnamefont {Brida}},\
  and\ \bibinfo {author} {\bibfnamefont {M.}~\bibnamefont {Genovese}},\
  }\bibfield  {title} {\bibinfo {title} {Experimental realization of quantum
  illumination},\ }\href {https://doi.org/10.1103/PhysRevLett.110.153603}
  {\bibfield  {journal} {\bibinfo  {journal} {Physical Review Letter}\ }\textbf
  {\bibinfo {volume} {110}},\ \bibinfo {pages} {153603} (\bibinfo {year}
  {2013})}\BibitemShut {NoStop}%
\bibitem [{\citenamefont {Zhang}\ \emph {et~al.}(2020)\citenamefont {Zhang},
  \citenamefont {England}, \citenamefont {Nomerotski}, \citenamefont {Svihra},
  \citenamefont {Ferrante}, \citenamefont {Hockett},\ and\ \citenamefont
  {Sussman}}]{Zhang_2020}%
  \BibitemOpen
  \bibfield  {author} {\bibinfo {author} {\bibfnamefont {Y.}~\bibnamefont
  {Zhang}}, \bibinfo {author} {\bibfnamefont {D.}~\bibnamefont {England}},
  \bibinfo {author} {\bibfnamefont {A.}~\bibnamefont {Nomerotski}}, \bibinfo
  {author} {\bibfnamefont {P.}~\bibnamefont {Svihra}}, \bibinfo {author}
  {\bibfnamefont {S.}~\bibnamefont {Ferrante}}, \bibinfo {author}
  {\bibfnamefont {P.}~\bibnamefont {Hockett}},\ and\ \bibinfo {author}
  {\bibfnamefont {B.}~\bibnamefont {Sussman}},\ }\bibfield  {title} {\bibinfo
  {title} {Multidimensional quantum-enhanced target detection via
  spectrotemporal-correlation measurements},\ }\href
  {https://doi.org/10.1103/PhysRevA.101.053808} {\bibfield  {journal} {\bibinfo
   {journal} {Phys. Rev. A}\ }\textbf {\bibinfo {volume} {101}},\ \bibinfo
  {pages} {053808} (\bibinfo {year} {2020})}\BibitemShut {NoStop}%
\bibitem [{\citenamefont {Gregory}\ \emph {et~al.}(2020)\citenamefont
  {Gregory}, \citenamefont {Moreau}, \citenamefont {Toninelli},\ and\
  \citenamefont {Padgett}}]{Gregory_2020}%
  \BibitemOpen
  \bibfield  {author} {\bibinfo {author} {\bibfnamefont {T.}~\bibnamefont
  {Gregory}}, \bibinfo {author} {\bibfnamefont {P.-A.}\ \bibnamefont {Moreau}},
  \bibinfo {author} {\bibfnamefont {E.}~\bibnamefont {Toninelli}},\ and\
  \bibinfo {author} {\bibfnamefont {M.~J.}\ \bibnamefont {Padgett}},\
  }\bibfield  {title} {\bibinfo {title} {Imaging through noise with quantum
  illumination},\ }\href {https://doi.org/10.1126/sciadv.aay2652} {\bibfield
  {journal} {\bibinfo  {journal} {Science Advances}\ }\textbf {\bibinfo
  {volume} {6}},\ \bibinfo {pages} {eaay2652} (\bibinfo {year} {2020})},\
  \Eprint
  {https://arxiv.org/abs/https://www.science.org/doi/pdf/10.1126/sciadv.aay2652}
  {https://www.science.org/doi/pdf/10.1126/sciadv.aay2652} \BibitemShut
  {NoStop}%
\bibitem [{\citenamefont {Pirandola}(2011)}]{Pirandola_2011}%
  \BibitemOpen
  \bibfield  {author} {\bibinfo {author} {\bibfnamefont {S.}~\bibnamefont
  {Pirandola}},\ }\bibfield  {title} {\bibinfo {title} {Quantum reading of a
  classical digital memory},\ }\href
  {https://doi.org/10.1103/PhysRevLett.106.090504} {\bibfield  {journal}
  {\bibinfo  {journal} {Physical Review Letter}\ }\textbf {\bibinfo {volume}
  {106}},\ \bibinfo {pages} {090504} (\bibinfo {year} {2011})}\BibitemShut
  {NoStop}%
\bibitem [{\citenamefont {Nair}(2011)}]{Nair_2011}%
  \BibitemOpen
  \bibfield  {author} {\bibinfo {author} {\bibfnamefont {R.}~\bibnamefont
  {Nair}},\ }\bibfield  {title} {\bibinfo {title} {Discriminating
  quantum-optical beam-splitter channels with number-diagonal signal states:
  Applications to quantum reading and target detection},\ }\href
  {https://doi.org/10.1103/PhysRevA.84.032312} {\bibfield  {journal} {\bibinfo
  {journal} {Physical Review A}\ }\textbf {\bibinfo {volume} {84}},\ \bibinfo
  {pages} {032312} (\bibinfo {year} {2011})}\BibitemShut {NoStop}%
\bibitem [{\citenamefont {DALL'ARNO}\ \emph {et~al.}(2012)\citenamefont
  {DALL'ARNO}, \citenamefont {BISIO},\ and\ \citenamefont
  {MAURO~D'ARIANO}}]{Dallarno_2012}%
  \BibitemOpen
  \bibfield  {author} {\bibinfo {author} {\bibfnamefont {M.}~\bibnamefont
  {DALL'ARNO}}, \bibinfo {author} {\bibfnamefont {A.}~\bibnamefont {BISIO}},\
  and\ \bibinfo {author} {\bibfnamefont {G.}~\bibnamefont {MAURO~D'ARIANO}},\
  }\bibfield  {title} {\bibinfo {title} {Ideal quantum reading of optical
  memories},\ }\href {https://doi.org/10.1142/S0219749912410109} {\bibfield
  {journal} {\bibinfo  {journal} {International Journal of Quantum
  Information}\ }\textbf {\bibinfo {volume} {10}},\ \bibinfo {pages} {1241010}
  (\bibinfo {year} {2012})},\ \Eprint
  {https://arxiv.org/abs/https://doi.org/10.1142/S0219749912410109}
  {https://doi.org/10.1142/S0219749912410109} \BibitemShut {NoStop}%
\bibitem [{\citenamefont {Wilde}\ \emph {et~al.}(2012)\citenamefont {Wilde},
  \citenamefont {Guha}, \citenamefont {Tan},\ and\ \citenamefont
  {Lloyd}}]{Wilde_2012}%
  \BibitemOpen
  \bibfield  {author} {\bibinfo {author} {\bibfnamefont {M.~M.}\ \bibnamefont
  {Wilde}}, \bibinfo {author} {\bibfnamefont {S.}~\bibnamefont {Guha}},
  \bibinfo {author} {\bibfnamefont {S.-H.}\ \bibnamefont {Tan}},\ and\ \bibinfo
  {author} {\bibfnamefont {S.}~\bibnamefont {Lloyd}},\ }\bibfield  {title}
  {\bibinfo {title} {Explicit capacity-achieving receivers for optical
  communication and quantum reading},\ }in\ \href
  {https://doi.org/10.1109/ISIT.2012.6284251} {\emph {\bibinfo {booktitle}
  {2012 IEEE International Symposium on Information Theory Proceedings}}}\
  (\bibinfo {year} {2012})\ pp.\ \bibinfo {pages} {551--555}\BibitemShut
  {NoStop}%
\bibitem [{\citenamefont {Dall'Arno}\ \emph {et~al.}(2012)\citenamefont
  {Dall'Arno}, \citenamefont {Bisio}, \citenamefont {D'Ariano}, \citenamefont
  {Mikov\'a}, \citenamefont {Je\ifmmode~\check{z}\else \v{z}\fi{}ek},\ and\
  \citenamefont {Du\ifmmode~\check{s}\else \v{s}\fi{}ek}}]{Dallarno_2012a}%
  \BibitemOpen
  \bibfield  {author} {\bibinfo {author} {\bibfnamefont {M.}~\bibnamefont
  {Dall'Arno}}, \bibinfo {author} {\bibfnamefont {A.}~\bibnamefont {Bisio}},
  \bibinfo {author} {\bibfnamefont {G.~M.}\ \bibnamefont {D'Ariano}}, \bibinfo
  {author} {\bibfnamefont {M.}~\bibnamefont {Mikov\'a}}, \bibinfo {author}
  {\bibfnamefont {M.}~\bibnamefont {Je\ifmmode~\check{z}\else \v{z}\fi{}ek}},\
  and\ \bibinfo {author} {\bibfnamefont {M.}~\bibnamefont
  {Du\ifmmode~\check{s}\else \v{s}\fi{}ek}},\ }\bibfield  {title} {\bibinfo
  {title} {Experimental implementation of unambiguous quantum reading},\ }\href
  {https://doi.org/10.1103/PhysRevA.85.012308} {\bibfield  {journal} {\bibinfo
  {journal} {Phys. Rev. A}\ }\textbf {\bibinfo {volume} {85}},\ \bibinfo
  {pages} {012308} (\bibinfo {year} {2012})}\BibitemShut {NoStop}%
\bibitem [{\citenamefont {Hirota}(2017)}]{Hirota_2017}%
  \BibitemOpen
  \bibfield  {author} {\bibinfo {author} {\bibfnamefont {O.}~\bibnamefont
  {Hirota}},\ }\bibfield  {title} {\bibinfo {title} {Error free quantum reading
  by quasi bell state of entangled coherent states},\ }\href
  {https://doi.org/https://doi.org/10.1515/qmetro-2017-0009} {\bibfield
  {journal} {\bibinfo  {journal} {Quantum Measurements and Quantum Metrology}\
  }\textbf {\bibinfo {volume} {4}},\ \bibinfo {pages} {70 } (\bibinfo {year}
  {2017})}\BibitemShut {NoStop}%
\bibitem [{\citenamefont {Fernandes~Pereira}\ and\ \citenamefont
  {Mancini}(2022)}]{Pereira_2022}%
  \BibitemOpen
  \bibfield  {author} {\bibinfo {author} {\bibfnamefont {F.~R.}\ \bibnamefont
  {Fernandes~Pereira}}\ and\ \bibinfo {author} {\bibfnamefont {S.}~\bibnamefont
  {Mancini}},\ }\bibfield  {title} {\bibinfo {title} {Error probability
  mitigation in quantum reading using classical codes},\ }\bibfield  {journal}
  {\bibinfo  {journal} {Entropy}\ }\textbf {\bibinfo {volume} {24}},\ \href
  {https://doi.org/10.3390/e24010005} {10.3390/e24010005} (\bibinfo {year}
  {2022})\BibitemShut {NoStop}%
\bibitem [{\citenamefont {Ortolano}\ \emph
  {et~al.}(2021{\natexlab{a}})\citenamefont {Ortolano}, \citenamefont {Losero},
  \citenamefont {Pirandola}, \citenamefont {Genovese},\ and\ \citenamefont
  {Ruo-Berchera}}]{Ortolano_2021}%
  \BibitemOpen
  \bibfield  {author} {\bibinfo {author} {\bibfnamefont {G.}~\bibnamefont
  {Ortolano}}, \bibinfo {author} {\bibfnamefont {E.}~\bibnamefont {Losero}},
  \bibinfo {author} {\bibfnamefont {S.}~\bibnamefont {Pirandola}}, \bibinfo
  {author} {\bibfnamefont {M.}~\bibnamefont {Genovese}},\ and\ \bibinfo
  {author} {\bibfnamefont {I.}~\bibnamefont {Ruo-Berchera}},\ }\bibfield
  {title} {\bibinfo {title} {Experimental quantum reading with photon
  counting},\ }\href@noop {} {\bibfield  {journal} {\bibinfo  {journal}
  {Science Advances}\ }\textbf {\bibinfo {volume} {7}},\ \bibinfo {pages}
  {eabc7796} (\bibinfo {year} {2021}{\natexlab{a}})}\BibitemShut {NoStop}%
\bibitem [{\citenamefont {Pirandola}\ \emph {et~al.}(2011)\citenamefont
  {Pirandola}, \citenamefont {Lupo}, \citenamefont {Giovannetti}, \citenamefont
  {Mancini},\ and\ \citenamefont {Braunstein}}]{Pirandola_2011a}%
  \BibitemOpen
  \bibfield  {author} {\bibinfo {author} {\bibfnamefont {S.}~\bibnamefont
  {Pirandola}}, \bibinfo {author} {\bibfnamefont {C.}~\bibnamefont {Lupo}},
  \bibinfo {author} {\bibfnamefont {V.}~\bibnamefont {Giovannetti}}, \bibinfo
  {author} {\bibfnamefont {S.}~\bibnamefont {Mancini}},\ and\ \bibinfo {author}
  {\bibfnamefont {S.~L.}\ \bibnamefont {Braunstein}},\ }\bibfield  {title}
  {\bibinfo {title} {Quantum reading capacity},\ }\href
  {https://doi.org/10.1088/1367-2630/13/11/113012} {\bibfield  {journal}
  {\bibinfo  {journal} {New Journal of Physics}\ }\textbf {\bibinfo {volume}
  {13}},\ \bibinfo {pages} {113012} (\bibinfo {year} {2011})}\BibitemShut
  {NoStop}%
\bibitem [{\citenamefont {Zhuang}\ and\ \citenamefont
  {Pirandola}(2020{\natexlab{a}})}]{Zhuang_2020}%
  \BibitemOpen
  \bibfield  {author} {\bibinfo {author} {\bibfnamefont {Q.}~\bibnamefont
  {Zhuang}}\ and\ \bibinfo {author} {\bibfnamefont {S.}~\bibnamefont
  {Pirandola}},\ }\bibfield  {title} {\bibinfo {title} {Entanglement-enhanced
  testing of multiple quantum hypotheses},\ }\href
  {https://doi.org/10.1038/s42005-020-0369-4} {\bibfield  {journal} {\bibinfo
  {journal} {Communications Physics}\ }\textbf {\bibinfo {volume} {3}},\
  \bibinfo {pages} {103} (\bibinfo {year} {2020}{\natexlab{a}})}\BibitemShut
  {NoStop}%
\bibitem [{\citenamefont {Oskouei}\ and\ \citenamefont
  {Mancini}(2021)}]{Oskuei_2021}%
  \BibitemOpen
  \bibfield  {author} {\bibinfo {author} {\bibfnamefont {S.~K.}\ \bibnamefont
  {Oskouei}}\ and\ \bibinfo {author} {\bibfnamefont {S.}~\bibnamefont
  {Mancini}},\ }\bibfield  {title} {\bibinfo {title} {Classical capacities of
  memoryless but not identical quantum channels},\ }\href
  {https://doi.org/10.1142/S0129055X21500124} {\bibfield  {journal} {\bibinfo
  {journal} {Reviews in Mathematical Physics}\ }\textbf {\bibinfo {volume}
  {33}},\ \bibinfo {pages} {2150012} (\bibinfo {year} {2021})},\ \Eprint
  {https://arxiv.org/abs/https://doi.org/10.1142/S0129055X21500124}
  {https://doi.org/10.1142/S0129055X21500124} \BibitemShut {NoStop}%
\bibitem [{\citenamefont {Revson}\ and\ \citenamefont
  {Mancini}(2021)}]{Revson_2021}%
  \BibitemOpen
  \bibfield  {author} {\bibinfo {author} {\bibfnamefont {F.}~\bibnamefont
  {Revson}}\ and\ \bibinfo {author} {\bibfnamefont {S.}~\bibnamefont
  {Mancini}},\ }\bibfield  {title} {\bibinfo {title} {Polar codes for quantum
  reading},\ }in\ \href {https://doi.org/10.1109/ISIT45174.2021.9517807} {\emph
  {\bibinfo {booktitle} {2021 IEEE International Symposium on Information
  Theory (ISIT)}}}\ (\bibinfo {year} {2021})\ pp.\ \bibinfo {pages}
  {2238--2243}\BibitemShut {NoStop}%
\bibitem [{\citenamefont {Ortolano}\ \emph
  {et~al.}(2021{\natexlab{b}})\citenamefont {Ortolano}, \citenamefont
  {Boucher}, \citenamefont {Degiovanni}, \citenamefont {Losero}, \citenamefont
  {Genovese},\ and\ \citenamefont {Ruo-Berchera}}]{Ortolano2021a}%
  \BibitemOpen
  \bibfield  {author} {\bibinfo {author} {\bibfnamefont {G.}~\bibnamefont
  {Ortolano}}, \bibinfo {author} {\bibfnamefont {P.}~\bibnamefont {Boucher}},
  \bibinfo {author} {\bibfnamefont {I.~P.}\ \bibnamefont {Degiovanni}},
  \bibinfo {author} {\bibfnamefont {E.}~\bibnamefont {Losero}}, \bibinfo
  {author} {\bibfnamefont {M.}~\bibnamefont {Genovese}},\ and\ \bibinfo
  {author} {\bibfnamefont {I.}~\bibnamefont {Ruo-Berchera}},\ }\bibfield
  {title} {\bibinfo {title} {Quantum conformance test},\ }\href
  {https://doi.org/10.1126/sciadv.abm3093} {\bibfield  {journal} {\bibinfo
  {journal} {Science Advances}\ }\textbf {\bibinfo {volume} {7}},\ \bibinfo
  {pages} {eabm3093} (\bibinfo {year} {2021}{\natexlab{b}})},\ \Eprint
  {https://arxiv.org/abs/https://www.science.org/doi/pdf/10.1126/sciadv.abm3093}
  {https://www.science.org/doi/pdf/10.1126/sciadv.abm3093} \BibitemShut
  {NoStop}%
\bibitem [{\citenamefont {Holevo}(1973)}]{Holevo_1973}%
  \BibitemOpen
  \bibfield  {author} {\bibinfo {author} {\bibfnamefont {A.~S.}\ \bibnamefont
  {Holevo}},\ }\bibfield  {title} {\bibinfo {title} {Bounds for the quantity of
  information transmitted by a quantum communication channel},\ }\href@noop {}
  {\bibfield  {journal} {\bibinfo  {journal} {Problems Inform. Transmission}\
  }\textbf {\bibinfo {volume} {9}},\ \bibinfo {pages} {177} (\bibinfo {year}
  {1973})}\BibitemShut {NoStop}%
\bibitem [{\citenamefont {Holevo}(1998)}]{Holevo_1998}%
  \BibitemOpen
  \bibfield  {author} {\bibinfo {author} {\bibfnamefont {A.}~\bibnamefont
  {Holevo}},\ }\bibfield  {title} {\bibinfo {title} {The capacity of the
  quantum channel with general signal states},\ }\href
  {https://doi.org/10.1109/18.651037} {\bibfield  {journal} {\bibinfo
  {journal} {IEEE Transactions on Information Theory}\ }\textbf {\bibinfo
  {volume} {44}},\ \bibinfo {pages} {269} (\bibinfo {year} {1998})}\BibitemShut
  {NoStop}%
\bibitem [{\citenamefont {Hausladen}\ \emph {et~al.}(1996)\citenamefont
  {Hausladen}, \citenamefont {Jozsa}, \citenamefont {Schumacher}, \citenamefont
  {Westmoreland},\ and\ \citenamefont {Wootters}}]{Hausladen_1996}%
  \BibitemOpen
  \bibfield  {author} {\bibinfo {author} {\bibfnamefont {P.}~\bibnamefont
  {Hausladen}}, \bibinfo {author} {\bibfnamefont {R.}~\bibnamefont {Jozsa}},
  \bibinfo {author} {\bibfnamefont {B.}~\bibnamefont {Schumacher}}, \bibinfo
  {author} {\bibfnamefont {M.}~\bibnamefont {Westmoreland}},\ and\ \bibinfo
  {author} {\bibfnamefont {W.~K.}\ \bibnamefont {Wootters}},\ }\bibfield
  {title} {\bibinfo {title} {Classical information capacity of a quantum
  channel},\ }\href {https://doi.org/10.1103/PhysRevA.54.1869} {\bibfield
  {journal} {\bibinfo  {journal} {Phys. Rev. A}\ }\textbf {\bibinfo {volume}
  {54}},\ \bibinfo {pages} {1869} (\bibinfo {year} {1996})}\BibitemShut
  {NoStop}%
\bibitem [{\citenamefont {Pirandola}\ \emph {et~al.}(2019)\citenamefont
  {Pirandola}, \citenamefont {Laurenza}, \citenamefont {Lupo},\ and\
  \citenamefont {Pereira}}]{Pirandola_2019}%
  \BibitemOpen
  \bibfield  {author} {\bibinfo {author} {\bibfnamefont {S.}~\bibnamefont
  {Pirandola}}, \bibinfo {author} {\bibfnamefont {R.}~\bibnamefont {Laurenza}},
  \bibinfo {author} {\bibfnamefont {C.}~\bibnamefont {Lupo}},\ and\ \bibinfo
  {author} {\bibfnamefont {J.~L.}\ \bibnamefont {Pereira}},\ }\bibfield
  {title} {\bibinfo {title} {Fundamental limits to quantum channel
  discrimination},\ }\href {https://doi.org/10.1038/s41534-019-0162-y}
  {\bibfield  {journal} {\bibinfo  {journal} {npj Quantum Information}\
  }\textbf {\bibinfo {volume} {5}},\ \bibinfo {pages} {50} (\bibinfo {year}
  {2019})}\BibitemShut {NoStop}%
\bibitem [{\citenamefont {Zhuang}\ and\ \citenamefont
  {Pirandola}(2020{\natexlab{b}})}]{Zhuang_2020a}%
  \BibitemOpen
  \bibfield  {author} {\bibinfo {author} {\bibfnamefont {Q.}~\bibnamefont
  {Zhuang}}\ and\ \bibinfo {author} {\bibfnamefont {S.}~\bibnamefont
  {Pirandola}},\ }\bibfield  {title} {\bibinfo {title} {Ultimate limits for
  multiple quantum channel discrimination},\ }\href
  {https://doi.org/10.1103/PhysRevLett.125.080505} {\bibfield  {journal}
  {\bibinfo  {journal} {Physical Review Letter}\ }\textbf {\bibinfo {volume}
  {125}},\ \bibinfo {pages} {080505} (\bibinfo {year}
  {2020}{\natexlab{b}})}\BibitemShut {NoStop}%
\bibitem [{\citenamefont {Holevo}(2012)}]{Holevo_2012}%
  \BibitemOpen
  \bibfield  {author} {\bibinfo {author} {\bibfnamefont {A.~S.}\ \bibnamefont
  {Holevo}},\ }\href {https://doi.org/doi:10.1515/9783110273403} {\emph
  {\bibinfo {title} {Quantum Systems, Channels, Information: A Mathematical
  Introduction}}}\ (\bibinfo  {publisher} {De Gruyter},\ \bibinfo {year}
  {2012})\BibitemShut {NoStop}%
\bibitem [{\citenamefont {Braunstein}\ and\ \citenamefont {van
  Loock}(2005)}]{Braunstein_2005}%
  \BibitemOpen
  \bibfield  {author} {\bibinfo {author} {\bibfnamefont {S.~L.}\ \bibnamefont
  {Braunstein}}\ and\ \bibinfo {author} {\bibfnamefont {P.}~\bibnamefont {van
  Loock}},\ }\bibfield  {title} {\bibinfo {title} {Quantum information with
  continuous variables},\ }\href {https://doi.org/10.1103/RevModPhys.77.513}
  {\bibfield  {journal} {\bibinfo  {journal} {Rev. Mod. Phys.}\ }\textbf
  {\bibinfo {volume} {77}},\ \bibinfo {pages} {513} (\bibinfo {year}
  {2005})}\BibitemShut {NoStop}%
\bibitem [{\citenamefont {Weedbrook}\ \emph {et~al.}(2012)\citenamefont
  {Weedbrook}, \citenamefont {Pirandola}, \citenamefont {Garc\'{\i}a-Patr\'on},
  \citenamefont {Cerf}, \citenamefont {Ralph}, \citenamefont {Shapiro},\ and\
  \citenamefont {Lloyd}}]{Weedbrook_2012}%
  \BibitemOpen
  \bibfield  {author} {\bibinfo {author} {\bibfnamefont {C.}~\bibnamefont
  {Weedbrook}}, \bibinfo {author} {\bibfnamefont {S.}~\bibnamefont
  {Pirandola}}, \bibinfo {author} {\bibfnamefont {R.}~\bibnamefont
  {Garc\'{\i}a-Patr\'on}}, \bibinfo {author} {\bibfnamefont {N.~J.}\
  \bibnamefont {Cerf}}, \bibinfo {author} {\bibfnamefont {T.~C.}\ \bibnamefont
  {Ralph}}, \bibinfo {author} {\bibfnamefont {J.~H.}\ \bibnamefont {Shapiro}},\
  and\ \bibinfo {author} {\bibfnamefont {S.}~\bibnamefont {Lloyd}},\ }\bibfield
   {title} {\bibinfo {title} {Gaussian quantum information},\ }\href
  {https://doi.org/10.1103/RevModPhys.84.621} {\bibfield  {journal} {\bibinfo
  {journal} {Reviews of Modern Physics}\ }\textbf {\bibinfo {volume} {84}},\
  \bibinfo {pages} {621} (\bibinfo {year} {2012})}\BibitemShut {NoStop}%
\bibitem [{\citenamefont {Bondani}\ \emph {et~al.}(2007)\citenamefont
  {Bondani}, \citenamefont {Allevi}, \citenamefont {Zambra}, \citenamefont
  {Paris},\ and\ \citenamefont {Andreoni}}]{Bondani_2007}%
  \BibitemOpen
  \bibfield  {author} {\bibinfo {author} {\bibfnamefont {M.}~\bibnamefont
  {Bondani}}, \bibinfo {author} {\bibfnamefont {A.}~\bibnamefont {Allevi}},
  \bibinfo {author} {\bibfnamefont {G.}~\bibnamefont {Zambra}}, \bibinfo
  {author} {\bibfnamefont {M.~G.~A.}\ \bibnamefont {Paris}},\ and\ \bibinfo
  {author} {\bibfnamefont {A.}~\bibnamefont {Andreoni}},\ }\bibfield  {title}
  {\bibinfo {title} {Sub-shot-noise photon-number correlation in a mesoscopic
  twin beam of light},\ }\href {https://doi.org/10.1103/PhysRevA.76.013833}
  {\bibfield  {journal} {\bibinfo  {journal} {Physical Review A}\ }\textbf
  {\bibinfo {volume} {76}},\ \bibinfo {pages} {013833} (\bibinfo {year}
  {2007})}\BibitemShut {NoStop}%
\bibitem [{\citenamefont {Avella}\ \emph {et~al.}(2016)\citenamefont {Avella},
  \citenamefont {Ruo-Berchera}, \citenamefont {Degiovanni}, \citenamefont
  {Brida},\ and\ \citenamefont {Genovese}}]{Avella_2016}%
  \BibitemOpen
  \bibfield  {author} {\bibinfo {author} {\bibfnamefont {A.}~\bibnamefont
  {Avella}}, \bibinfo {author} {\bibfnamefont {I.}~\bibnamefont
  {Ruo-Berchera}}, \bibinfo {author} {\bibfnamefont {I.~P.}\ \bibnamefont
  {Degiovanni}}, \bibinfo {author} {\bibfnamefont {G.}~\bibnamefont {Brida}},\
  and\ \bibinfo {author} {\bibfnamefont {M.}~\bibnamefont {Genovese}},\
  }\bibfield  {title} {\bibinfo {title} {Absolute calibration of an emccd
  camera by quantum correlation, linking photon counting to the analog
  regime},\ }\href {https://doi.org/10.1364/OL.41.001841} {\bibfield  {journal}
  {\bibinfo  {journal} {Optics Letter}\ }\textbf {\bibinfo {volume} {41}},\
  \bibinfo {pages} {1841} (\bibinfo {year} {2016})}\BibitemShut {NoStop}%
\bibitem [{\citenamefont {Meda}\ \emph {et~al.}(2017)\citenamefont {Meda},
  \citenamefont {Losero}, \citenamefont {Samantaray}, \citenamefont
  {Scafirimuto}, \citenamefont {Pradyumna}, \citenamefont {Avella},
  \citenamefont {Ruo-Berchera},\ and\ \citenamefont {Genovese}}]{Meda_2017}%
  \BibitemOpen
  \bibfield  {author} {\bibinfo {author} {\bibfnamefont {A.}~\bibnamefont
  {Meda}}, \bibinfo {author} {\bibfnamefont {E.}~\bibnamefont {Losero}},
  \bibinfo {author} {\bibfnamefont {N.}~\bibnamefont {Samantaray}}, \bibinfo
  {author} {\bibfnamefont {F.}~\bibnamefont {Scafirimuto}}, \bibinfo {author}
  {\bibfnamefont {S.}~\bibnamefont {Pradyumna}}, \bibinfo {author}
  {\bibfnamefont {A.}~\bibnamefont {Avella}}, \bibinfo {author} {\bibfnamefont
  {I.}~\bibnamefont {Ruo-Berchera}},\ and\ \bibinfo {author} {\bibfnamefont
  {M.}~\bibnamefont {Genovese}},\ }\bibfield  {title} {\bibinfo {title}
  {Photon-number correlation for quantum enhanced imaging and sensing},\ }\href
  {https://doi.org/10.1088/2040-8986/aa7b27} {\bibfield  {journal} {\bibinfo
  {journal} {Journal of Optics}\ }\textbf {\bibinfo {volume} {19}},\ \bibinfo
  {pages} {094002} (\bibinfo {year} {2017})}\BibitemShut {NoStop}%
\bibitem [{\citenamefont {Nielsen}\ and\ \citenamefont
  {Chuang}(2011)}]{Nielsen_2011}%
  \BibitemOpen
  \bibfield  {author} {\bibinfo {author} {\bibfnamefont {M.~A.}\ \bibnamefont
  {Nielsen}}\ and\ \bibinfo {author} {\bibfnamefont {I.~L.}\ \bibnamefont
  {Chuang}},\ }\href@noop {} {\emph {\bibinfo {title} {Quantum Computation and
  Quantum Information}}},\ \bibinfo {edition} {10th}\ ed.\ (\bibinfo
  {publisher} {Cambridge University Press},\ \bibinfo {address} {USA},\
  \bibinfo {year} {2011})\BibitemShut {NoStop}%
\bibitem [{\citenamefont {Shi}\ \emph {et~al.}(2020)\citenamefont {Shi},
  \citenamefont {Zhang}, \citenamefont {Pirandola},\ and\ \citenamefont
  {Zhuang}}]{Shi_2020}%
  \BibitemOpen
  \bibfield  {author} {\bibinfo {author} {\bibfnamefont {H.}~\bibnamefont
  {Shi}}, \bibinfo {author} {\bibfnamefont {Z.}~\bibnamefont {Zhang}}, \bibinfo
  {author} {\bibfnamefont {S.}~\bibnamefont {Pirandola}},\ and\ \bibinfo
  {author} {\bibfnamefont {Q.}~\bibnamefont {Zhuang}},\ }\bibfield  {title}
  {\bibinfo {title} {Entanglement-assisted absorption spectroscopy},\ }\href
  {https://doi.org/10.1103/PhysRevLett.125.180502} {\bibfield  {journal}
  {\bibinfo  {journal} {Phys. Rev. Lett.}\ }\textbf {\bibinfo {volume} {125}},\
  \bibinfo {pages} {180502} (\bibinfo {year} {2020})}\BibitemShut {NoStop}%
\bibitem [{\citenamefont {Gagatsos}\ \emph {et~al.}(2016)\citenamefont
  {Gagatsos}, \citenamefont {Karanikas}, \citenamefont {Kordas},\ and\
  \citenamefont {Cerf}}]{Gagatsos_2016}%
  \BibitemOpen
  \bibfield  {author} {\bibinfo {author} {\bibfnamefont {C.~N.}\ \bibnamefont
  {Gagatsos}}, \bibinfo {author} {\bibfnamefont {A.~I.}\ \bibnamefont
  {Karanikas}}, \bibinfo {author} {\bibfnamefont {G.}~\bibnamefont {Kordas}},\
  and\ \bibinfo {author} {\bibfnamefont {N.~J.}\ \bibnamefont {Cerf}},\
  }\bibfield  {title} {\bibinfo {title} {Entropy generation in gaussian quantum
  transformations: applying the replica method to continuous-variable quantum
  information theory},\ }\href {https://doi.org/10.1038/npjqi.2015.8}
  {\bibfield  {journal} {\bibinfo  {journal} {npj Quantum Information}\
  }\textbf {\bibinfo {volume} {2}},\ \bibinfo {pages} {15008} (\bibinfo {year}
  {2016})}\BibitemShut {NoStop}%
\end{thebibliography}%
\end{document}